\documentclass[acmsmall,screen]{acmart}

\setcopyright{rightsretained}
\acmPrice{}
\acmDOI{10.1145/3563347}
\acmYear{2022}
\copyrightyear{2022}
\acmSubmissionID{oopslab22main-p566-p}
\acmJournal{PACMPL}
\acmVolume{6}
\acmNumber{OOPSLA2}
\acmArticle{184}
\acmMonth{10}

\usepackage{booktabs}   
\usepackage{subcaption} 
\usepackage{algorithmicx}
\usepackage{algpseudocode}
\usepackage{algorithm}
\usepackage{wrapfig}
\usepackage{tikz}
\usepackage{amsthm}
\usepackage{mathpartir}
\usepackage{xspace}
\usepackage{graphicx}
\usepackage{pict2e}
\usepackage{pifont}
\usepackage{xspace}
\usepackage{listings}
\usepackage{enumitem}
\usepackage{letltxmacro}
\usepackage{caption}
\usepackage{subcaption}
\newcommand*{\SavedLstInline}{}
\LetLtxMacro\SavedLstInline\lstinline
\DeclareRobustCommand*{\lstinline}{%
  \ifmmode
    \let\SavedBGroup\bgroup
    \def\bgroup{%
      \let\bgroup\SavedBGroup
      \hbox\bgroup
    }%
  \fi
  \SavedLstInline
}
\lstdefinelanguage{zelus}
   {morekeywords={
	let,in,rec,where,end,if,then,else,do,done,run,
	open,
	fun,node,hybrid,proba,const,
	match,with,automaton,emit,
	pre,when,whenot,fby,merge,on,clock,
	or,and,not,as,mod,
	unless,until,continue,reset,every,await,
	init,der,type,last,
	period,local,present,sample, observe, factor, infer, eval, value
    },
    sensitive=true,
    morecomment=[n]{(*}{*)},
    morestring=[b]",
    escapechar=\%,
    columns=fullflexible,
    keepspaces=true,
    basicstyle=\ttfamily,
    mathescape=true,
    }

\def\zl{\lstinline[basicstyle=\small\ttfamily]}

\newcommand{\probzelusexpr}{\mathit{e}}
\newcommand{\ProbzelusExpr}{\mathit{Expr}}

\newcommand{\ProbzelusEq}{\mathit{Eq}}

\newcommand{\Node}{\mathit{Decl}}

\newcommand{\Distr}{D}
\newcommand{\distr}{d}

\newcommand{\Expr}{E}
\newcommand{\expr}{e}

\newcommand{\Real}{{\mathbb{R}}}
\newcommand{\real}{{r}}
\newcommand{\Val}{\mathbb{V}}

\newcommand{\Integer}{{\mathbb{Z}}}
\newcommand{\integer}{c}

\newcommand{\Randomvar}{{\mathcal{V}}}
\newcommand{\randomvar}{{X}}

\newcommand{\normal}[2]{{\mathcal{N}({#1}, {#2})}}
\newcommand{\normalbig}[2]{{\mathcal{N}\left({#1}, {#2}\right)}}
\newcommand{\bern}[1]{{\mathrm{Bernoulli}({#1})}}
\newcommand{\bernbig}[1]{{\mathrm{Bernoulli}\left({#1}\right)}}
\newcommand{\betad}[2]{{\beta({#1}, {#2})}}
\newcommand{\deltad}[1]{{\delta({#1})}}

\newcommand{\plus}[2]{{{#1} \; \texttt{+} \; {#2}}}
\newcommand{\plusbold}[2]{{{#1} \oplus {#2}}}
\newcommand{\minus}[2]{{{#1} \; \texttt{-} \; {#2}}}
\newcommand{\mult}[2]{{{#1} \; \texttt{*} \; {#2}}}
\newcommand{\division}[2]{\frac{{#1}}{{#2}}}
\newcommand{\squareroot}[1]{{\sqrt{{#1}}}}
\newcommand{\summation}[3]{{\texttt{sum} \; {#1} \; \texttt{in} \; {#2} \; \texttt{:} \; {#3}}}
\newcommand{\ite}[3]{{\texttt{ite(} \; {#1}, {#2}, {#3} \; \texttt{)}}}

\newcommand{\bnfarrow}{\Coloneqq}

\newcommand{\pstate}{g}
\newcommand{\Pstate}{G}
\newcommand{\pstateempty}{\ensuremath{\{\}}}

\newcommand{\sem}[1]{{\llbracket {#1} \rrbracket}}
\newcommand{\semi}[1]{\sem{#1}^{\text{init}}}
\newcommand{\sems}[2]{\sem{#1}_{#2}^{\text{step}}}
\newcommand{\psem}[1]{\{\mkern-3.8mu[ #1 ]\mkern-3.8mu\}}
\newcommand{\psemi}[1]{\psem{#1}^{\text{init}}}
\newcommand{\psems}[2]{\psem{#1}_{#2}^{\text{step}}}
\newcommand{\letin}[1]{\ensuremath{\mathit{let} \; #1 \; \textit{in} \;}}

\newcommand{\env}{\gamma}

\newcommand{\mem}{m}
\newcommand{\Mem}{M}

\DeclareMathOperator{\bind}{\gg\!\!=}

\usepackage{transparent}
\usepackage{xcolor}
\definecolor{pfcolor}{RGB}{110,180,227}
\newcommand{\pfbullet}{{\color{pfcolor}{\ensuremath{\blacksquare}}}\xspace}
\definecolor{bdscolor}{RGB}{229,158,12}

\definecolor{pmdscolor}{RGB}{147,3,211}
\newcommand{\pmdsbullet}{{\color{pmdscolor}{\ensuremath{\blacktriangledown}}}\xspace}
\definecolor{ndscolor}{RGB}{69,155,118}

\definecolor{ssdscolor}{RGB}{238,226,97}
\newcommand{\ssdsbullet}{{\color{ssdscolor}{\ensuremath{\blacklozenge}}}\xspace}

\newcommand{\gcdsacro}{DS\xspace}
\newcommand{\ssdsacro}{SSI\xspace}

\newtheorem{axiom}{Axiom}

\algblock[TryCatchFinally]{try}{endtry}
\algcblockdefx[TryCatchFinally]{TryCatchFinally}{catch}{endtry}
  [1]{\textbf{catch} #1}
  {\textbf{end try}}

\begin{document}

\title{Semi-symbolic Inference for Efficient Streaming Probabilistic Programming}         


\author{Eric Atkinson}
\affiliation{
  \institution{MIT}            
  \country{USA}                    
}

\author{Charles Yuan}
\affiliation{
  \institution{MIT}            
  \country{USA}                    
}

\author{Guillaume Baudart}
\affiliation{
  \institution{ENS -- PSL University -- CNRS -- Inria}           
  \department{DI ENS}
  \country{France}                   
}

\author{Louis Mandel}
\affiliation{
  \institution{IBM Research}            
  \department{MIT-IBM Watson AI Lab}
  \country{USA}                    
}

\author{Michael Carbin}
\affiliation{
  \institution{MIT}            
  \country{USA}                    
}
\begin{abstract}
A streaming probabilistic program receives a stream of observations and produces a stream of distributions that are conditioned on these observations.
Efficient inference is often possible in a streaming context using Rao-Blackwellized particle filters (RBPFs), which exactly solve inference problems when possible and fall back on sampling approximations when necessary.
While RBPFs can be implemented by hand to provide efficient inference, the goal of streaming probabilistic programming is to automatically generate such efficient inference implementations given input probabilistic programs.

In this work, we propose \emph{semi-symbolic inference}, a technique for executing probabilistic programs using a runtime inference system that automatically implements Rao-Blackwellized particle filtering.
To perform exact and approximate inference together, the semi-symbolic inference system manipulates symbolic distributions to perform exact inference when possible and falls back on approximate sampling when necessary.
This approach enables the system to implement the same RBPF a developer would write by hand.
To ensure this, we identify \emph{closed families} of distributions -- such as linear-Gaussian and finite discrete models -- on which the inference system guarantees exact inference.
We have implemented the runtime inference system in the ProbZelus streaming probabilistic programming language.
Despite an average $1.6\times$ slowdown compared to the state of the art on existing benchmarks, our evaluation shows that speedups of $3\times$--$87\times$ are obtainable on a new set of challenging benchmarks we have designed to exploit closed families.

\end{abstract}


%
%
%

\maketitle

\section{Introduction}

\emph{Probabilistic programming languages} enable developers to describe a probabilistic model in a programming language and let the language's compiler and runtime perform Bayesian inference~\citep{goodman_stuhlmuller_2014,TranHSB0B17,pyro,MurrayS18,tolpin_et_al_2016}.
In this work, we focus on \emph{streaming} probabilistic programs, as first formalized in~\citet{rppl}.
In a streaming probabilistic program, the program receives a stream of observations and produces a stream of distributions that are conditioned on these observations.

\paragraph{Streaming Inference}
Rao-Blackwellized particle filtering~\citep{rbpf} is a state-of-the-art inference technique that can be used in a streaming context.
Rao-Blackwellized particle filters (RBPFs) exactly solve inference problems when possible~(i.e., when a closed-form solution exists) and fall back on sampling-based approximate particle filtering~\cite{pf} when symbolic computations fail.
The key challenge to applying RBPFs is designing an effective state representation that maintains both a sample-based representation and a symbolic representation, with as much of the state as possible in the symbolic representation.

\paragraph{Semi-Symbolic Inference}
In this work, we propose \emph{semi-symbolic inference}, in which a particle filter is augmented with a symbolic state consisting of mathematical expressions that encode distributions of random variables in the program.
At runtime, the semi-symbolic inference system transforms the expressions in the symbolic state according to closed-form solutions from probability theory.
Compared to previous work using a different state representation~\citep{rppl}, semi-symbolic inference can maintain an exact representation in more cases.
In particular, we prove that the semi-symbolic inference system guarantees an exact representation on \emph{closed families}, which include linear-Gaussian and finite discrete probabilistic models.

\paragraph{Contributions}

In this paper, we present the following contributions:
\begin{itemize}
  \item We present semi-symbolic inference, a new technique for Rao-Blackwellized particle filtering in streaming probabilistic programs. In Section~\ref{sec:semisymb}, we define the state representation as well as the operations the semi-symbolic inference system uses to perform inference.
  \item We discuss the guarantees and advantages of semi-symbolic inference. In Section~\ref{sec:properties}, we state and prove several theorems about closed families. In particular, we show that the semi-symbolic inference system provides guaranteed exact inference on linear-Gaussian and finite discrete probabilistic models, which developers can use to reliably write models that the inference system implements as RBPFs.
  \item We implement semi-symbolic inference inside the streaming probabilistic programming language ProbZelus~\citep{rppl}. The implementation is available at \url{https://github.com/ibm/probzelus}, and has been accepted as an artifact~\cite{artifact}.
  \item We evaluate semi-symbolic inference in ProbZelus on a set of benchmarks in Section~\ref{sec:evaluation}, and compare against prior work on \emph{delayed sampling}~\citep{murray_ds} in ProbZelus.
    We show that semi-symbolic inference has a slowdown of $1.6\times$ on existing benchmarks. 
    In exchange for this overhead, by maintaining exact representations more often than delayed sampling, semi-symbolic inference can achieve speedups of $3\times$--$87\times$ on a set of challenging benchmarks that exercise the closed family guarantee.
    We discuss in detail two of these benchmarks that illustrate the situations in which delayed sampling fails to perform exact inference but semi-symbolic inference can.
\end{itemize}
By executing streaming probabilistic programs with semi-symbolic inference, developers can combine the efficiency of exact inference with the generality of approximate inference.
The closed family guarantee further ensures that the programs will deliver the performance developers expect.
This paper is an extension of our OOPSLA 2022 paper~\cite{semisymb_oopsla22} with appendices.

\section{Example}
\label{sec:example}

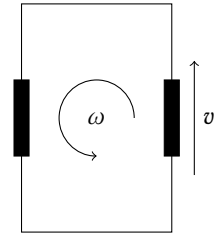
\begin{wrapfigure}{R}{0.3\textwidth}
\vspace*{-3em}
\begin{center}
\begin{tikzpicture}
\draw (0, 0) rectangle (2, 3);
\draw [fill=black] (-0.1, 2) rectangle (0.1, 1);
\draw [fill=black] (1.9, 2) rectangle (2.1, 1);
\draw [->] (1.5, 1.5) arc (0:270:0.5);
\draw (1, 1.5) node {$\omega$};
\draw [->] (2.3, 0.75) -- (2.3, 2.25);
\draw (2.5, 1.5) node {$v$};
\end{tikzpicture}
\end{center}
\caption{Diagram of a two-wheeled robot. The objective is to estimate the angular velocity $\omega$ and the forward velocity $v$ using wheel sensors.}
\label{fig:diagram}
\end{wrapfigure}

To demonstrate semi-symbolic inference, we use the streaming probabilistic programming language ProbZelus to model a robot with two wheels.
Figure~\ref{fig:diagram} shows a diagram depicting the robot.
Our objective is to estimate the angular and forward velocity of the robot using sensors that measure the speed of each wheel.
Such a task is often a precursor to estimating the position of the robot, as described, for example, in~\citet{wheels_paper}.

In this section, we first explain how to implement such a model in the ProbZelus streaming probabilistic programming language.
We then explain how our implementation of a semi-symbolic runtime inference system for ProbZelus executes this program.

\begin{figure}
\begin{lstlisting}[basicstyle=\small\ttfamily,aboveskip=0em,numbers=left]
let proba wheels (left_rate, right_rate) = (vel, omega) where%\label{code:proba}%
  rec init omega = 0. and init vel = 0.%\label{code:init}%
  and omega = sample (gaussian (last omega, omega_var))%\label{code:omega}%
  and vel = sample (gaussian (last vel, vel_var))%\label{code:vel}%
  and () = observe (gaussian (vel -. wb *. omega, sensor_err), left_rate)%\label{code:obsl}%
  and () = observe (gaussian (vel +. wb *. omega, sensor_err), right_rate)%\label{code:obsr}%
\end{lstlisting}
\caption{A program that encodes the robot model in ProbZelus. The model inputs are two streams \lstinline{left_rate} and \lstinline{right_rate} encoding the speed sensors on each wheel, and the outputs are streams of the estimated velocity \lstinline{vel} and angular velocity \lstinline{omega}. The constants \lstinline{omega_var}, \lstinline{vel_var}, and \lstinline{sensor_err} specify variances of \lstinline{omega} and \lstinline{vel}, and the wheel sensors respectively, and \lstinline{wb} specifies the width of the robot's wheel base. In our example, we assume values $\lstinline{omega_var} = 2500$, $\lstinline{vel_var} = 2500$, $\lstinline{wb} = 2$, and $\lstinline{sensor_err} = 1$.}
\label{fig:wheels_code}
\end{figure}

\subsection{Implementation in ProbZelus}

Figure~\ref{fig:wheels_code} presents an implementation of the model in the ProbZelus streaming probabilistic programming language.
The core objects in ProbZelus are \emph{stream functions} that transform input streams into output streams.
The \zl$proba$ keyword on Line~\ref{code:proba} signifies the definition of a stream function whose definition may include probabilistic operators.
The remainder of the line specifies that the \zl$wheels$ stream function takes as input a pair of streams \zl$left_rate$ and \zl$right_rate$ encoding the speed sensors on each wheel, and returns a pair of the robot's estimated velocity \zl$vel$ and angular velocity \zl$omega$. These estimated velocities are defined by the subsequent mutually recursive equations.

Lines~\ref{code:init} and~\ref{code:omega} specify a probabilistic model for the angular velocity \zl$omega$.
At each time step, Line~\ref{code:omega} specifies that \zl$omega$ is a stream of values, each sampled from a Gaussian distribution using the \zl$sample$ probabilistic operator.
The mean of the Gaussian is given by the previous value in the stream of \zl$omega$ values as specified using the \zl$last omega$ syntax, except that at the first time step, \zl$last omega$ takes on the initial value of 0 as specified on Line~\ref{code:init} by the \zl$init$ keyword.
The variance of the Gaussian is given by the constant \zl$omega_var$.

Lines~\ref{code:init} and~\ref{code:vel} specify a model for the forward velocity \zl$vel$ symmetric to that for angular velocity.

Line~\ref{code:obsl} conditions the model by observing the left wheel's velocity from the forward velocity and angular velocity.
In particular, it specifies that the left wheel's velocity is sampled from a Gaussian distribution.
The mean of this distribution subtracts from the forward velocity \zl$vel$ the angular velocity \zl$omega$ multiplied by the constant \zl$wb$, which represents the width of the robot's wheel base.
The program then specifies, using the \zl$observe$ probabilistic operator, that the model is conditioned on this Gaussian random variable being equal to the input value \zl$left_rate$.
Line~\ref{code:obsr} specifies a similar observation to for the right wheel, except that the product of \zl$omega$ and \zl$wb$ is added to \zl$vel$ instead of subtracted.
For both observations, the Gaussian's variance is the constant \zl$sensor_err$.

\subsection{Semi-Symbolic Inference}

\begin{figure}
\begin{subfigure}[t]{0.48\textwidth}
\tikzstyle{var}=[circle, draw=black, fill=gray!30, minimum width=0.2cm]
\begin{center}
\begin{tikzpicture}
\draw (0, 0.8) node (veltxt) {\texttt{vel}};
\draw (0, 0) node [draw=black, fill=gray!30, circle] (velnode) {$\randomvar_v$};
\draw[->, densely dotted] (veltxt) -- (velnode);
\draw (-1.5, 0) node (veldistr) {$\normal{0}{2500}$};
\draw[->, densely dotted] (velnode) to [bend right=45] (veldistr) ;

\draw (1, 0.75) node (omegatxt) {\texttt{omega}};
\draw (1, 0) node [draw=black, fill=gray!30, circle] (omeganode) {$\randomvar_o$};
\draw[->, densely dotted] (omegatxt) -- (omeganode);
\draw (2.5, 0) node (omegadistr) {$\normal{0}{2500}$};
\draw[->, densely dotted] (omeganode) to [bend left=45] (omegadistr) ;
\end{tikzpicture}
\end{center}
\caption{The symbolic state after executing Lines~\ref{code:omega} and~\ref{code:vel}. Each program variable points to a new random variable with a Gaussian distribution.}
\label{fig:postsample}
\end{subfigure}$\quad$
\begin{subfigure}[t]{0.48\textwidth}
\tikzstyle{var}=[circle, draw=black, fill=gray!30, minimum width=0.2cm]
\begin{tikzpicture}
\draw (0, 0.8) node (veltxt) {\texttt{vel}};
\draw (0, 0) node [draw=black, fill=gray!30, circle] (velnode) {$\randomvar_v$};
\draw[->, densely dotted] (veltxt) -- (velnode);
\draw (-1.5, 0) node (veldistr) {$\normal{0}{2500}$};
\draw[->, densely dotted] (velnode) to [bend right=45] (veldistr) ;

\draw (1, 0.75) node (omegatxt) {\texttt{omega}};
\draw (1, 0) node [draw=black, fill=gray!30, circle] (omeganode) {$\randomvar_o$};
\draw[->, densely dotted] (omegatxt) -- (omeganode);
\draw (2.5, 0) node (omegadistr) {$\normal{0}{2500}$};
\draw[->, densely dotted] (omeganode) to [bend left=45] (omegadistr) ;

\draw (0, -1) node [draw=black, fill=gray!30, circle] (leftnode) {$\randomvar_l$};
\draw[->] (velnode) -- (leftnode);
\draw[->] (omeganode) -- (leftnode);
\draw (-2, -1) node (leftdistr) {$\normal{\minus{\randomvar_v}{\mult{2}{\randomvar_o}}}{1}$};
\draw[->, densely dotted] (leftnode) to [bend left=45] (leftdistr) ;
\end{tikzpicture}
\caption{The first step in executing the \zl$observe$ on Line~\ref{code:obsl} adds a new random variable with the appropriate distribution. The symbolic distribution contains references to both previous random variables. }
\label{fig:obsdiag1}
\end{subfigure}\\
\begin{subfigure}[t]{\textwidth}
\tikzstyle{var}=[circle, draw=black, fill=gray!30, minimum width=0.2cm]
\begin{center}
\begin{tikzpicture}
\draw (0, 0.8) node (veltxt) {\texttt{vel}};
\draw (0, 0) node [draw=black, fill=gray!30, circle] (velnode) {$\randomvar_v$};
\draw[->, densely dotted] (veltxt) -- (velnode);
\draw (-4.5, 0) node (veldistr) {$\normalbig{\plus{\left(\mult{\left(\plus{\division{\mult{-2}{\randomvar_o}}{2500}}{\division{\randomvar_l}{1}}\right)}{\division{1}{\plus{\division{1}{2500}}{\division{1}{1}}}}\right)}{(\mult{2}{\randomvar_o})}}{\division{1}{\plus{\division{1}{2500}}{\division{1}{1}}}}$};
\draw[->, densely dotted] (velnode) to [bend right=45] (veldistr) ;

\draw (1, 0.75) node (omegatxt) {\texttt{omega}};
\draw (1, 0) node [draw=black, fill=gray!30, circle] (omeganode) {$\randomvar_o$};
\draw[->, densely dotted] (omegatxt) -- (omeganode);
\draw (2.5, 0) node (omegadistr) {$\normal{0}{2500}$};
\draw[->, densely dotted] (omeganode) to [bend left=45] (omegadistr) ;

\draw (0, -1) node [draw=black, fill=gray!30, circle] (leftnode) {$\randomvar_l$};
\draw[->, densely dotted] (leftnode) to [bend left=45] (leftdistr) ;
\draw[->] (leftnode) -- (velnode);
\draw[->] (omeganode) -- (leftnode);
\draw[->] (omeganode) -- (velnode);
\draw (-2.2, -1) node (leftdistr) {$\normal{\minus{0}{\mult{2}{\randomvar_o}}}{\plus{2500}{1}}$};
\end{tikzpicture}
\end{center}
\vspace{-0.2cm}
\caption{Next, the inference system swaps the new random variable $\randomvar_l$ and $\randomvar_v$.
The result is that $\randomvar_l$ no longer depends on $\randomvar_v$ and $\randomvar_v$ depends on $\randomvar_l$ and $\randomvar_o$.}
\label{fig:obsdiag2}
\end{subfigure}\\
\begin{subfigure}[b]{\textwidth}
\tikzstyle{var}=[circle, draw=black, fill=gray!30, minimum width=0.2cm]
\begin{tikzpicture}
\draw (0, 0.8) node (veltxt) {\texttt{vel}};
\draw (0, 0) node [draw=black, fill=gray!30, circle] (velnode) {$\randomvar_v$};
\draw[->, densely dotted] (veltxt) -- (velnode);
\draw (-4.5, 0) node (veldistr) {$\normalbig{\plus{\left(\mult{\left(\plus{\division{\mult{-2}{\randomvar_o}}{2500}}{\division{\randomvar_l}{1}}\right)}{\division{1}{\plus{\division{1}{2500}}{\division{1}{1}}}}\right)}{(\mult{2}{\randomvar_o})}}{\division{1}{\plus{\division{1}{2500}}{\division{1}{1}}}}$};
\draw[->, densely dotted] (velnode) to [bend right=45] (veldistr) ;

\draw (1, 0.75) node (omegatxt) {\texttt{omega}};
\draw (1, 0) node [draw=black, fill=gray!30, circle] (omeganode) {$\randomvar_o$};
\draw[->, densely dotted] (omegatxt) -- (omeganode);
\draw (1.5, -2.0) node (omegadistr) {$\normalbig{\division{\mult{\left(\plus{\division{\mult{-2}{0}}{\mult{\mult{-2}{-2}}{2500}}}{\division{\randomvar_l}{\plus{2500}{1}}}\right)}{\division{1}{\plus{\division{1}{\mult{\mult{-2}{-2}}{2500}}}{\division{1}{\plus{2500}{1}}}}}}{-2}}{ \division{  \division{1}{\plus{\division{1}{\mult{\mult{-2}{-2}}{2500}}}{\division{1}{\plus{2500}{1}}}} }{\mult{-2}{-2}}  }$};
\draw[->, densely dotted] (omeganode) to [bend left=45] (omegadistr) ;

\draw (0, -1) node [draw=black, fill=gray!30, circle] (leftnode) {$\randomvar_l$};
\draw[->, densely dotted] (leftnode) to [bend left=25] (leftdistr) ;
\draw[->] (leftnode) -- (velnode);
\draw[->] (leftnode) -- (omeganode);
\draw[->] (omeganode) -- (velnode);
\draw (-4.0, -1) node (leftdistr) {$\normal{\minus{0}{\mult{2}{0}}}{\plus{(\mult{\mult{2}{2}}{2500})}{(\plus{2500}{1})}}$};
\end{tikzpicture}
\caption{Next, the inference system swaps $\randomvar_l$ with $\randomvar_o$.
The distribution of $\randomvar_l$ now has no dependencies.}
\label{fig:obsdiag3}
\end{subfigure}\\
\begin{subfigure}[b]{\textwidth}
\tikzstyle{var}=[circle, draw=black, fill=gray!30, minimum width=0.2cm]
\begin{tikzpicture}
\draw (0, 0.8) node (veltxt) {\texttt{vel}};
\draw (0, 0) node [draw=black, fill=gray!30, circle] (velnode) {$\randomvar_v$};
\draw[->, densely dotted] (veltxt) -- (velnode);
\draw (-4.5, 0) node (veldistr) {$\normalbig{\plus{\left(\mult{\left(\plus{\division{\mult{-2}{\randomvar_o}}{2500}}{\division{\randomvar_l}{1}}\right)}{\division{1}{\plus{\division{1}{2500}}{\division{1}{1}}}}\right)}{(\mult{2}{\randomvar_o})}}{\division{1}{\plus{\division{1}{2500}}{\division{1}{1}}}}$};
\draw[->, densely dotted] (velnode) to [bend right=45] (veldistr) ;

\draw (1, 0.75) node (omegatxt) {\texttt{omega}};
\draw (1, 0) node [draw=black, fill=gray!30, circle] (omeganode) {$\randomvar_o$};
\draw[->, densely dotted] (omegatxt) -- (omeganode);
\draw (1.5, -2.0) node (omegadistr) {$\normalbig{\division{\mult{\left(\plus{\division{\mult{-2}{0}}{\mult{\mult{-2}{-2}}{2500}}}{\division{\randomvar_l}{\plus{2500}{1}}}\right)}{\division{1}{\plus{\division{1}{\mult{\mult{-2}{-2}}{2500}}}{\division{1}{\plus{2500}{1}}}}}}{-2}}{ \division{  \division{1}{\plus{\division{1}{\mult{\mult{-2}{-2}}{2500}}}{\division{1}{\plus{2500}{1}}}} }{\mult{-2}{-2}}  }$};
\draw[->, densely dotted] (omeganode) to [bend left=45] (omegadistr) ;

\draw (0, -1) node [draw=black, fill=gray!30, circle] (leftnode) {$\randomvar_l$};
\draw[->, densely dotted] (leftnode) to [bend left=15] (leftdistr) ;
\draw[->] (leftnode) -- (velnode);
\draw[->] (leftnode) -- (omeganode);
\draw[->] (omeganode) -- (velnode);
\draw (-4.0, -1) node [xshift=20pt](leftdistr) {$\deltad{-1}$};
\end{tikzpicture}
\caption{Then, the system intervenes to replace $\randomvar_l$'s distribution with a Delta distribution.}
\label{fig:intervenediag}
\end{subfigure}

\caption{Depiction of the evolution of the symbolic state during the execution of the program in Figure~\ref{fig:wheels_code} under semi-symbolic inference.
We use gray circles represent random variables, and solid arrows to represent conceptual dependencies between random variables.
Dotted lines point a) from program variables to the random variables they refer to, and b) from random variables to their symbolic distributions.
We also depict the current values of the \zl$vel$ and \zl$omega$ program variables.}
\label{fig:symb_diag}
\end{figure}
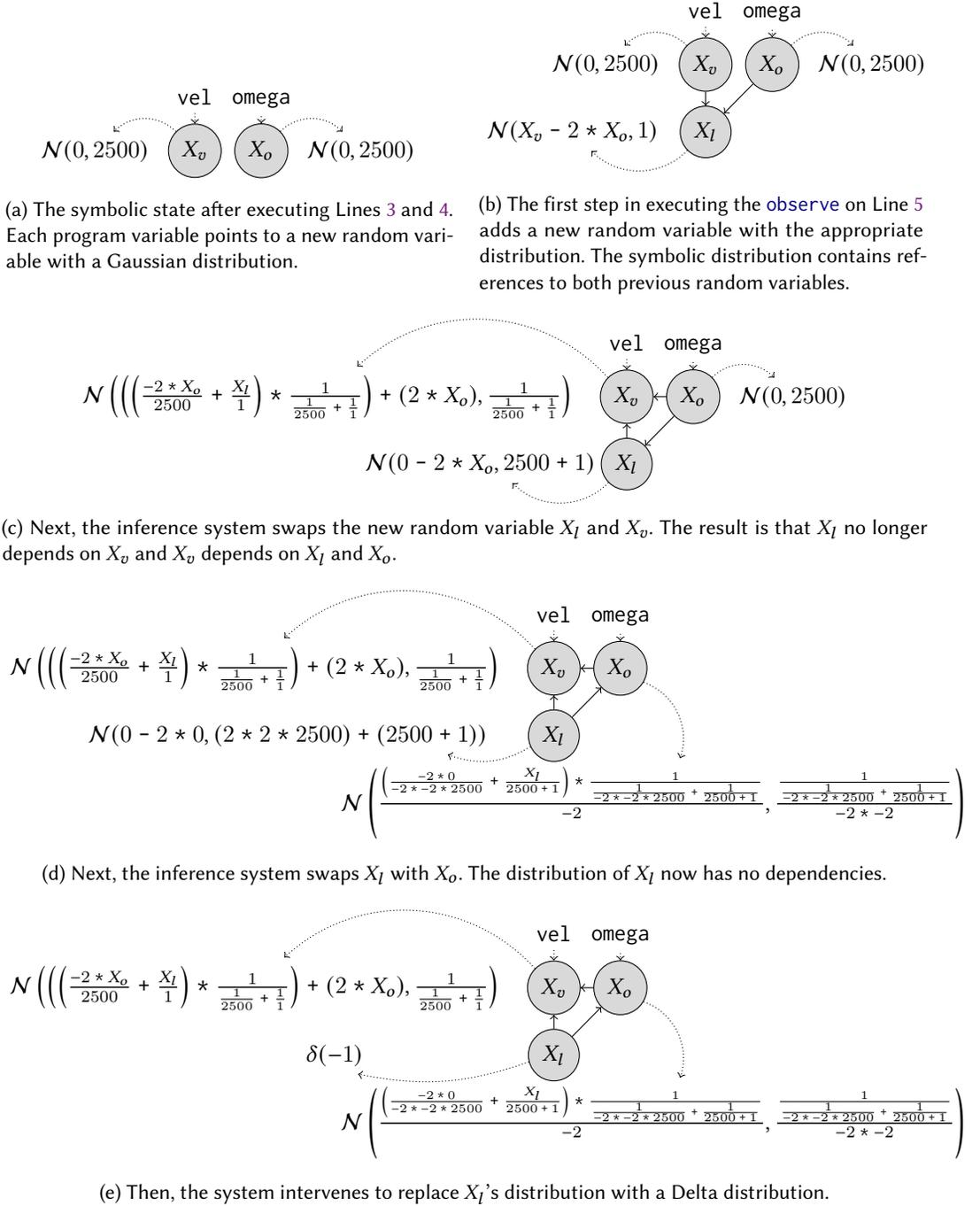

We now describe how to use semi-symbolic inference to execute the ProbZelus program in Figure~\ref{fig:wheels_code}.
In particular, we describe how the \emph{symbolic state} of the semi-symbolic runtime inference system evolves over the course of the first iteration of the \zl$wheels$ stream function.
The symbolic states described in this section are depicted in Figure~\ref{fig:symb_diag}.

\paragraph{Sampling}

First, the semi-symbolic inference system executes the \zl$sample$ operators on Lines~\ref{code:omega} and~\ref{code:vel}.
The semi-symbolic implementation of \zl$sample$ constructs symbolic terms representing distributions and returns the random variables $\randomvar_v$ and $\randomvar_o$ that point to these distributions.
It stores handles to $\randomvar_v$ and $\randomvar_o$ inside the program variables \zl$vel$ and \zl$omega$, respectively.

Figure~\ref{fig:postsample} depicts the symbolic state after both of these \zl$sample$ operations.
In these depictions, we assume the constant variances \zl$omega_var$ and \zl$vel_var$ are both equal to $2500$, corresponding to a standard deviation of $\sqrt{2500} = 50$ for both forward and angular velocity.

\paragraph{Observation}

Next, the semi-symbolic inference system executes the \zl$observe$ operation on Line~\ref{code:obsl}.
Like in standard probabilistic languages (e.g. \citet{goodman_stuhlmuller_2014}), the \zl$observe$ operation performs a scoring operation.
However, in semi-symbolic inference, it also has to update the symbolic state.
Here, we describe the symbolic state update.
The inference system first constructs a
random variable $\randomvar_l$ representing the distribution specified on Line~\ref{code:obsl}.
Where the program refers to the variables \zl$vel$ and \zl$omega$, in the corresponding symbolic distribution these expressions contain handles to the corresponding random variables $\randomvar_v$ and $\randomvar_o$.

Figure~\ref{fig:obsdiag1} depicts the resulting symbolic state.
In these depictions, we assume that the constant variance \zl$sample_err$ is $1$ and the constant \zl$wb$ is 2.

\paragraph{Swapping}

To perform the observation, the semi-symbolic inference system first converts this new random variable into a \emph{root}, meaning a random variable with no parents. A parent of a random variable is another random variable that is mentioned in any of its sub-expressions.
To perform this conversion, the system performs a \emph{swap}, the core operation of semi-symbolic inference.
A swap changes the dependency order between two random variables in the symbolic state.
The swap ensures that the overall joint distribution of variables in the symbolic state does not change.

In the example, the inference system tries to swap
$\randomvar_l$ with the random variable to which \zl$vel$ points, $\randomvar_v$.
The system must first determine if such a swap is possible.
Based on the state depicted in Figure~\ref{fig:obsdiag1}, the system detects that
$\randomvar_l$ can be written as an affine function of $\randomvar_v$, namely $\randomvar_l = a * \randomvar_v + b$ where $a = 1$ and $b = \mult{-2}{\randomvar_o}$ (note that random variables other than $\randomvar_v$ may appear in $a$ and $b$).
Because the two distributions are Gaussians with constant variance, and they are related by an affine function, the system determines that a swap is possible.

\paragraph{Conjugate Priors}

The mathematical principle behind a swap is that of \emph{conjugate priors}~\citep{conjugate_priors}.
Conjugate priors provide rules for manipulating distributions that are useful for performing exact inference.
The rules change the direction of dependencies while preserving the overall joint distribution between the involved random variables.
While an advanced understanding of the theory of conjugate priors is not necessary to understand semi-symbolic inference, we briefly explain how conjugate priors apply to the swap of $\randomvar_l$ and $\randomvar_v$ in the example.
This swap uses the conjugate rule that given random variables $\randomvar_c \sim \normal{\mu_c}{\sigma_c^2}$ and $\randomvar_d \sim \normal{\randomvar_c}{\sigma_d^2}$, then the distribution of $\randomvar_d$ is equivalent to $\normal{\mu_c}{\sigma_c^2 + \sigma_d^2}$, and the distribution of $\randomvar_c$ is equivalent to $$\randomvar_c \sim \normalbig{\left(\tfrac{\mu_c}{\sigma_c^2} + \tfrac{\randomvar_d}{\sigma_d^2}\right) * \sigma_\mathrm{cond}^2}{ \sigma_\mathrm{cond}^2} \text{, where } \sigma_\mathrm{cond}^2 = \left(\tfrac{1}{\sigma_c^2} + \tfrac{1}{\sigma_d ^2}\right)^{-1}$$
The swap also uses the general rule of linear transformations of Gaussians: if $\randomvar_a \sim \normal{\mu_a}{\sigma_a^2}$, and $\randomvar_b = \plus{\mult{a}{\randomvar_a}}{b}$, then $\randomvar_b \sim \normal{\plus{\mult{a}{\mu_a}}{b}}{\mult{a^2}{\sigma_a^2}}$.

Figure~\ref{fig:obsdiag2} depicts the resulting state of the swap of $\randomvar_l$ and $\randomvar_v$.
The distribution of the variable to which \zl$omega$ points is unchanged, but the other two random variables have new distributions to satisfy their reversed dependence relationship while maintaining the correct joint distribution.

Next, the inference system swaps the variable to which \zl$omega$ points, $\randomvar_o$, with
$\randomvar_l$.
The system determines that these variables have an affine relationship, namely $\randomvar_l = a * \randomvar_o + b$ where $a = -2$ and $b = 0$, and can therefore be swapped.
Figure~\ref{fig:obsdiag3} depicts the results.
The system replaces the distribution of $\randomvar_l$ and $\randomvar_o$, and leaves the distribution of $\randomvar_v$ unchanged.
Note that at this point, the distribution of $\randomvar_l$ does not depend on any other random variables, and can be evaluated to produce a closed-form distribution.

\paragraph{Intervention}

Next, the inference system performs the observation by intervening on the value of~$\randomvar_l$.
Let us assume that the input observed value (i.e., the value of~\zl$left_rate$) is $-1$.
To condition the symbolic state on the fact that $\randomvar_l$ is equal to $-1$, the system replaces the distribution of $\randomvar_l$ with the \emph{Delta distribution} $\deltad{-1}$, the distribution with all mass at $-1$.
The result is depicted in Figure~\ref{fig:intervenediag}.

\paragraph{Simplification}

The system further simplifies the symbolic state in Figure~\ref{fig:intervenediag} using the fact that a random variable sampled from a Delta distribution can be replaced with the interior expression of the Delta. In this example, this means that when the variable
$\randomvar_l$ appears in a distribution expression, the system can replace it with $-1$. Using this fact and subsequent evaluation steps, the distribution for $\randomvar_o$ evaluates to $\normal{0.4}{500}$, and the distribution for $\randomvar_v$ partially evaluates to $\normalbig{\left(\frac{-2 * X_o}{2500} + -1\right) * 1}{1}$. To facilitate this simplification, the semi-symbolic inference system incorporates a partial evaluator, which we describe in Section~\ref{sec:partial-evaluator}.

\paragraph{Iteration}

The inference system performs the observation on Line~\ref{code:obsr} using a similar sequence of steps.
This concludes the system's operations at the first iteration.
Subsequent iterations execute similar operations, except that \zl$vel$ and \zl$omega$ are sampled by reference to their previous value instead of their initial value 0.
In subsequent symbolic states, the random variables to which \zl$vel$ and \zl$omega$ point contain symbolic expressions that refer to variables sampled at earlier iterations.

\paragraph{Accuracy}
As a result of the symbolic manipulations, the distributions of \zl$vel$ and \zl$omega$ become closed-form symbolic expressions. These expressions estimate the forward and angular velocity of the robot and may then be used to estimate additional properties such as its position. Importantly, the distributions are exact, avoiding any loss of accuracy introduced by sampling approximations.

This is a notable improvement over the prior implementation of ProbZelus, which implemented inference using \emph{delayed sampling}~\cite{murray_ds}.
Delayed sampling executes similarly to semi-symbolic inference, but has a different symbolic state representation.
Delayed sampling cannot represent the distributions of \zl$vel$ and \zl$omega$ exactly, and thus falls back on approximate sampling.
We further discuss the implementation of ProbZelus in Section~\ref{sec:background}, and provide more details on the specific differences between delayed sampling and semi-symbolic inference Section~\ref{sec:ds_comparison}.

\section{Background: ProbZelus Syntax and Semantics}
\label{sec:background}

In this section, we review from \citet{rppl} the syntax and semantics of ProbZelus, the streaming probabilistic programming language within which we implement semi-symbolic inference.
We consider a fragment of ProbZelus to illustrate the core concepts of how the system works. A full presentation of the semantics reviewed here can be found in~\citet{rppl}.

\begin{figure}
\begin{align*}
\Node \bnfarrow \; & \zl{let proba} \; f \; x \; \zl{=} \; \ProbzelusExpr \;
\mid \; \Node \; \Node\\
\ProbzelusExpr \bnfarrow \; & x \; \mid \; c \; \mid \; \zl{(} \ProbzelusExpr\zl{,}\,\ProbzelusExpr\zl{)} \; \mid \; \mathit{op}(\ProbzelusExpr) \; \mid \; f(\ProbzelusExpr) \; \mid \; \zl{last} \; x \; \mid \; \ProbzelusExpr \; \zl$where rec$ \; \ProbzelusEq\\
\mid \; &  \zl{present} \; \ProbzelusExpr \;\; \zl{->} \; \ProbzelusExpr \; \zl{else} \; \ProbzelusExpr \; \mid \; \zl{reset} \; \ProbzelusExpr \; \zl{every} \; \ProbzelusExpr \\
\ProbzelusEq \bnfarrow \; & x = \ProbzelusExpr \; \mid \; \zl{init} \; x = c \; \mid \; \ProbzelusEq \; \zl{and} \; \ProbzelusEq
\end{align*}
\caption{Syntax for a fragment of the ProbZelus language.}
\label{fig:syntax}
\end{figure}

\paragraph{Syntax}

We describe the fragment of ProbZelus presented in Figure~\ref{fig:syntax}.
A program is given by a sequence of stream function declarations each introduced by the keyword \zl{proba}.

An expression is a variable $x$, a constant $c$, a pair of expressions, an operator application $\mathit{op}$, a function application $f$, an access to the previous value of a variable with \zl$last$, or an expression whose free variables are defined by a set of mutually recursive equations given by \zl$where rec$.
The \zl$present$ construct executes one of its two branches depending on the value of the first expression which can be a Boolean expression or a sporadic signal. The \zl$reset$ expression re-initializes the state of the first expression (i.e.\ \zl{last}~$x$ returns $x$'s inital value) each time the second expression is true.

The contents of a \zl$where rec$ definition either assigns a variable to an expression or specifies an initial value for a variable using the \zl$init$ keyword.
The initial value works in conjunction with the \zl$last$ expression to provide the value \zl$last$ returns on the first time step.

The operators of the language include standard arithmetic operators as well as the probabilistic operators \zl{sample} to draw from a distribution and \zl{observe} to condition on an observation. The operator \zl{value} forces the inference system to explicitly draw a sample. The purpose of the \zl{value} operator is to enable developers to explicitly control the location of random sampling.

\paragraph{Semantic Model}

The semantics of a deterministic expression~$\probzelusexpr$ is given by a pair of an initial memory state $\semi{\probzelusexpr} \in \Mem$ and a step function $\sems{\probzelusexpr}{\env} \in \Mem \mapsto O \times \Mem$.
The step function is parameterized by an environment~$\env$ that contains the value of the free variables of~$\probzelusexpr$.
The step function takes the current memory and outputs a result value and an updated memory.

A program produces a stream of outputs ${(o_j)}_{j \in \mathbb{N}}$ by repeatedly executing the step function on the stream of inputs contained in the environment~$\gamma_i$ starting from the initial memory~$m_0$:
\begin{align*}
\mem_0 = \; & \semi{\probzelusexpr}\\
o_1, \mem_1 = \; & \sems{\probzelusexpr}{\env_1}(m_0)\\
o_2, \mem_2 = \; & \sems{\probzelusexpr}{\env_2}(m_1)\\
\dots\\
o_n, \mem_n = \; & \sems{\probzelusexpr}{\env_n}(m_{n-1})\\
\dots
\end{align*}

In contrast, the step function of a probabilistic expression takes the current memory and returns a measure over pairs (result, new memory): $\psems{\probzelusexpr}{\env} \in \Mem \mapsto \Sigma_{O \times \Mem} \mapsto [0, \infty)$ where $\Sigma_{O \times \Mem}$ denotes the $\sigma$-algebra over pairs of results and memory.
This measure is then normalized and split into a distribution of results and a distribution of memories.
At each step, we integrate the step function over the current distribution of memories to compute the next distribution of results and the next distribution of memories.
Details can be found in~\cite[Section~3.3]{rppl}.

\subsection{Particle Filtering Semantics}

\begin{figure}
$\begin{array}{rcl}
\psemi{c}\phantom{\,(\mem, w)} & = & \zl*()*
\\
\psems{c}{\env}(\mem, w) & = & (c, \mem, w)
\\[.5em]
\psemi{x}\phantom{\,(\mem, w)} & = & \zl*()*
\\
\psems{x}{\env}(\mem, w) & = & (\env(x), \mem, w)
\\[.5em]
\psemi{\zl*last* \; x}\phantom{\,(\mem, w)} & = & \zl*()*
\\
\psems{\zl*last* \; x}{\env}(\mem, w) & = & (\env (x\_\zl*last*), \mem, w)
\\[.5em]
\psemi{\mathit{op} \zl*(* \probzelusexpr \zl*)*}\phantom{\,(\mem, w)} & = &
\psemi{\probzelusexpr}
\\
\psems{\mathit{op} \zl*(* \probzelusexpr \zl*)*}{\env}(\mem, w) & = &
\letin{v, \mem', w' = \psems{\probzelusexpr}{\env}(\mem,w)} \\ &&
(\mathit{op}(v), \mem', w')
\\[.5em]
\psemi{\zl*sample(*\probzelusexpr\zl*)*}\phantom{\,(\mem, w)} & = &
  \psemi{\probzelusexpr}
\\
\psems{\zl*sample(*\probzelusexpr\zl*)*}{\env}(\mem, w) & = &
  \letin{\mu, \mem', w' = \psems{\probzelusexpr}{\env}(\mem,w)} \\ &&
  (\textsc{draw}(\mu), \mem', w')
\\[.5em]
\psemi{\zl*observe(*\probzelusexpr_\mu\zl*,*\probzelusexpr_v\zl*)*}\phantom{\,((\mem_\mu, \mem_v), w)} & = &
  (\psemi{\probzelusexpr_\mu}, \psemi{\probzelusexpr_v})
\\
\psems{\zl*observe(*\probzelusexpr_\mu\zl*,*\probzelusexpr_v\zl*)*}{\env}((\mem_\mu, \mem_v), w) & = &
  \letin{\mu, \mem_\mu', w_1 = \psems{\probzelusexpr_\mu}{\env}(\mem_\mu, w)} \\ &&
  \letin{v, \mem_v', w_2 = \psems{\probzelusexpr_v}{\env}(\mem_v, w_1)} \\ &&
  (\zl*()*, (\mem_\mu', \mem_v'), w_2 * \textsc{score}(\mu, v))
\\[.5em]
\psemi{\zl*value(*\probzelusexpr\zl*)*}\phantom{\,(\mem, w)} & = & \psemi{\probzelusexpr}
\\
\psems{\zl*value(*\probzelusexpr\zl*)*}{\env}(\mem, w) & = &
  \psems{\probzelusexpr}{\env}(\mem, w)
\\[.5em]
\multicolumn{3}{l}{
\left \{ \mkern-5mu \left [
\begin{array}{@{}r@{~}l@{}}
\probzelusexpr \; \zl*where rec*
 & \zl*init* \; x_1 = c_1 \; \zl*and* \; \dots \; \zl*and init* \; x_k = c_k\\
 & \zl*and* \; y_1 = \probzelusexpr_1 \; \zl*and* \; \dots \; \zl*and* \; y_n = \probzelusexpr_n
\end{array}
\right ] \mkern-5mu \right \}^{\text{init}}
}\\
& = &
 ((c_1, \dots, c_k), (\psemi{\probzelusexpr_1}, \dots, \psemi{\probzelusexpr_n}),\psemi{\probzelusexpr})
\\
\multicolumn{3}{l}{
\left \{ \mkern-5mu \left [
\begin{array}{@{}r@{~}l@{}}
\probzelusexpr \; \zl*where rec*
 & \zl*init* \; x_1 = c_1 \; \zl*and* \; \dots \; \zl*and init* \; x_k = c_k\\
 & \zl*and* \; y_1 = \probzelusexpr_1 \; \zl*and* \; \dots \; \zl*and* \; y_n = \probzelusexpr_n
\end{array}
\right ] \mkern-5mu \right \}^{\text{step}}_{\env}
(((v_1, \dots, v_k),(\mem_1, \dots, \mem_n), m), w)
}\\
& = &
  \letin{\env_1 = \env[v_1/x_1\_\zl*last*]} \\ &&
  \dots \\ &&
  \letin{\env_k = \env[v_k/x_k\_\zl*last*]} \\ &&
  \letin{v_1, \mem_1', w_1 = \psems{\probzelusexpr_1}{\env_k}(\mem_1, w)}\\ &&
  \letin{\env_1' = \env_k[y_1/v_1]}\\ &&
  \dots \\ &&
  \letin{v_{n}, \mem_{n}', w_{n} = \psems{\probzelusexpr_n}{\env_{n-1}'}(\mem_{n}, w_{n-1})}\\ &&
  \letin{\env_n' = \env_{n-1}'[y_{n}/v_{n}]}\\ &&
  \letin{v, \mem', w' = \psems{\probzelusexpr}{\env_n'}(\mem, w_n)}\\ &&
  (v, ((\env_n'(x_1), \dots, \env_n'(x_k)), (\mem_1', \dots, \mem_n'), \mem'), w')
\end{array}$
\caption{ Particle filtering semantics for a subset of ProbZelus constructs.  }
\label{fig:pf_sem}
\end{figure}

The semantics of an expression~$\probzelusexpr$ may be given by approximate sampling through \emph{particle filtering}.
Particle filtering defines the semantics of~$\probzelusexpr$ by performing multiple independent executions~(particles) of~$\probzelusexpr$ and tracking the likelihood of each execution.
We define the semantics of each particle for an expression $\probzelusexpr$ by an initial memory $\psemi{\probzelusexpr}$ and a step function $\psems{\probzelusexpr}{\env}$.
The step function of each particle takes a memory~$\mem$ and a real-valued weight~$w$ as input, and returns the value of the expression, the updated memory for the next iteration, and the updated weight.

Figure~\ref{fig:pf_sem} presents the particle filtering semantics of selected ProbZelus expressions.
The initial memory of a constant is empty and is represented with the value \zl{()}.
Its step function returns the value of the constant and leaves the memory and weight unchanged.
Similarly, an access to a variable or the last value of a variable does not update the memory, and its value is taken from the environment~$\env$.
The semantics of the application of an operator $\mathit{op}(\probzelusexpr)$, e.g. $+$ or $*$,
applies the operator to the evaluation of the sub-expression $\probzelusexpr$ and propagates the memory and the weight.

The expression $\probzelusexpr~\zl*where*~\zl*rec*~\ProbzelusEq$ introduces and updates the state variables $x_1, \dots, x_k$, which must be the subset of the variables $y_1, \dots, y_n$ used as \zl{last $y_i$}.
To define the semantics, we assume that all \zl{init} equations appear first and other equations are sorted according to their data dependencies where the \zl{last} operator does not introduce a dependency.
The memory is composed of a slot for each variable introduced by an \zl{init} equation and the memory required for each sub-expression.
The step function puts all the state variables in the environment~$\env_k$ and then evaluates each equation to compute the current value of each variable.
Finally, the expression returns the value of $\probzelusexpr$ evaluated in the environment containing the value of all the variables. The memory is updated with the current value of the state variables.

The semantics of \zl{sample(}$\probzelusexpr$\zl{)} introduces probability into program execution.
The step function evaluates the sub-expression~$\probzelusexpr$ to obtain a distribution~$\mu$ and draws a random value from~$\mu$ using the \textsc{draw} primitive.
The memory of this expression is the memory of the sub-expression.

The expression \zl{observe(}$\probzelusexpr_\mu, \probzelusexpr_v$\zl{)} conditions the model using the weight~$w$.
The step function evaluates the two sub-expressions to obtain a distribution~$\mu$ and a value~$v$ and updates the weight by multiplying it by $\textsc{score}(\mu, v)$, the value of the probability density of~$\mu$ in~$v$, i.e. $\textsc{score}(\mu, v) = \mu_{\text{pdf}}(v)$.

In the particle filter semantics, the \zl*value* function simply evaluates its argument.

\paragraph{Inference.} In this work, we present inference as a transformation called \zl{infer} operating on ProbZelus expressions that can contain free variables whose values are defined in the environment~$\env$.
The semantics of \zl{infer} on expression~$\probzelusexpr$ defines a stream of distributions as an initial memory and a step function.
The memory is the distribution of possible memory configurations for~$\probzelusexpr$.
The step function computes the distribution of outputs and the distribution of memories by performing~$N$ independent executions of~$e$ with the input memory sampled from the distribution of memories from the previous iteration.

For particle filtering~(PF), the \zl{infer} construct is defined as follows:
\[
\begin{array}{rcl}
\semi{\zl*infer*_{\text{PF}}(N, e)}\phantom{\,(\mem)} & = &
  \delta_{\psemi{e}}
\\[.5em]
\sems{\zl*infer*_{\text{PF}}(N, e)}{\env}(\mem) & = &
  \begin{array}[t]{@{}l}
    \letin{\left[
      \begin{array}{@{}l}
      v_i, \mem_i', w_i' =
        \begin{array}[t]{@{}l}
        \letin{\mem_i = \textsc{draw}(\mem)}\\
        \psems{e}{\env}(\mem_i, 1)
        \end{array}
      \end{array}\hspace{-1em}
    \right]_{1 \le i \le N}} \\
    \letin{\mu = \lambda U.\ \Sigma_{i=1}^N \overline{w_i} * \delta_{(v_i, m_i)}(U)}\\
    (\pi_{1*}(\mu), \pi_{2*}(\mu))
  \end{array}
\end{array}
\]

The initial memory is a Delta distribution on the initial memory of~$\probzelusexpr$.
The step function first builds a size-$N$ array containing the result of~$N$ executions of~$\probzelusexpr$ on memories randomly drawn from the input memory distribution.
Then, the distribution~$\mu$ of pairs of outputs and memory is computed using the weight from each particle: $\overline{w_i} = w_i' / \Sigma_{j=1}^N w_j'$. Finally, the individual distributions of outputs and memories are separated using the pushforward of $\mu$ across the projections $\pi_1$ and $\pi_2$.

\subsection{Delayed Sampling Semantics}
\label{sec:semisymb_sec}

\newcommand{\reveal}[1]{\texttransparent{1.0}{#1}}
\newcommand{\mask}[1]{\texttransparent{0.4}{#1}}

\begin{figure}
\mask{
$\begin{array}{rcl}
\psemi{c}\phantom{\,(\mem, \pstate, w)} & = & \zl*()*
\\
\psems{c}{\env}(\mem, \reveal{\pstate}, w) & = & (c, \mem, \reveal{\pstate}, w)
\\[.5em]
\psemi{x}\phantom{\,(\mem, \pstate, w)} & = & \zl*()*
\\
\psems{x}{\env}(\mem, \reveal{\pstate}, w) & = & (\env(x), \mem, \reveal{\pstate}, w)
\\[.5em]
\psemi{\zl*last* \; x}\phantom{\,(\mem, \pstate, w)} & = & \zl*()*
\\
\psems{\zl*last* \; x}{\env}(\mem, \reveal{\pstate}, w) & = & (\env (x\_\zl*last*), \mem, \reveal{\pstate}, w)
\\[.5em]
\psemi{\mathit{op} \zl*(* \probzelusexpr \zl*)*}\phantom{\,(\mem, w)} & = &
\psemi{\probzelusexpr}
\\
\psems{\mathit{op} \zl*(* \probzelusexpr \zl*)*}{\env}(\mem, \reveal{\pstate}, w) & = &
\letin{v, \mem', \reveal{\pstate'}, w' = \psems{\probzelusexpr}{\env}(\mem, \reveal{\pstate}, w)} \\ &&
(\reveal{\texttt{app}(\mathit{op},v)}, \mem', \reveal{\pstate'}, w')
\\[.5em]
\psemi{\zl*sample(*\probzelusexpr\zl*)*}\phantom{\,(\mem, \pstate, w)} & = &
  \psemi{\probzelusexpr}
\\
\psems{\zl*sample(*\probzelusexpr\zl*)*}{\env}(\mem, \reveal{\pstate}, w) & = &
  \letin{\mu, \mem', \reveal{\pstate'}, w' = \psems{\probzelusexpr}{\env}(\mem, \reveal{\pstate}, w)} \\ &&
  \reveal{\letin{X, \pstate'' = \textsc{assume}(\mu, \pstate')}}\\ &&
  (\reveal{X}, \mem', \reveal{\pstate''}, w')
\\[.5em]
\psemi{\zl*observe(*\probzelusexpr_\mu\zl*,*\probzelusexpr_v\zl*)*}\phantom{\,((\mem_\mu, \mem_v), \pstate, w)} & = &
  (\psemi{\probzelusexpr_\mu}, \psemi{\probzelusexpr_v})
\\
\psems{\zl*observe(*\probzelusexpr_\mu\zl*,*\probzelusexpr_v\zl*)*}{\env}((\mem_\mu, \mem_v), \reveal{\pstate}, w) & = &
  \letin{\mu, \mem_\mu', \reveal{\pstate_\mu}, w_1 = \psems{\probzelusexpr_\mu}{\env}(\mem_\mu, \reveal{\pstate}, w)} \\ &&
  \reveal{\letin{X, \pstate_x = \textsc{assume}(\mu, \pstate_\mu)}} \\ &&
  \letin{v, \mem_v', \reveal{\pstate_v}, w_2 = \psems{\probzelusexpr_v}{\env}(\mem_v, \reveal{\pstate_x}, w_1)} \\ &&
  \reveal{\letin{v', g_v' = \textsc{value}(v, g_v)}}\\ &&
  \reveal{\letin{\pstate', s = \textsc{observe}(X, v', \pstate_v')}} \\ &&
  (\zl*()*, (\mem_\mu', \mem_v'), \reveal{\pstate'}, \reveal{s * w_2})
\\[.5em]
\psemi{\zl*value(*\probzelusexpr\zl*)*}\phantom{(\mem, \pstate, w)} & = &
\psemi{\probzelusexpr}
\\
\psems{\zl*value(*\probzelusexpr\zl*)*}{\env}(\mem, \reveal{\pstate}, w) & = &
\reveal{\letin{\randomvar, \mem', \pstate', w' = \psems{e}{\env}(\mem, \pstate, w)}}\\ &&
  \reveal{\letin{v, \pstate'' = \textsc{value}(\randomvar, \pstate')}}\\ &&
  \reveal{(v, \mem', \pstate'', w')}
\\[.5em]
\multicolumn{3}{l}{
\left \{ \mkern-5mu \left [
\begin{array}{@{}r@{~}l@{}}
\probzelusexpr \; \zl*where rec*
 & \zl*init* \; x_1 = c_1 \; \zl*and* \; \dots \; \zl*and init* \; x_k = c_k\\
 & \zl*and* \; y_1 = \probzelusexpr_1 \; \zl*and* \; \dots \; \zl*and* \; y_n = \probzelusexpr_n
\end{array}
\right ] \mkern-5mu \right \}^{\text{init}}
}\\
& = &
 ((c_1, \dots, c_k), (\psemi{\probzelusexpr_1}, \dots, \psemi{\probzelusexpr_n}),\psemi{\probzelusexpr})
\\
\multicolumn{3}{l}{
\left \{ \mkern-5mu \left [
\begin{array}{@{}r@{~}l@{}}
\probzelusexpr \; \zl*where rec*
 & \zl*init* \; x_1 = c_1 \; \zl*and* \; \dots \; \zl*and init* \; x_k = c_k\\
 & \zl*and* \; y_1 = \probzelusexpr_1 \; \zl*and* \; \dots \; \zl*and* \; y_n = \probzelusexpr_n
\end{array}
\right ] \mkern-5mu \right \}^{\text{step}}_{\env}
(((v_1, \dots, v_k),(\mem_1, \dots, \mem_n), m), \reveal{\pstate}, w)
}\\
& = &
  \letin{\env_1 = \env[v_1/x_1\_\zl*last*]} \\ &&
  \dots \\ &&
  \letin{\env_k = \env[v_k/x_k\_\zl*last*]} \\ &&
  \letin{v_1, \mem_1', \reveal{\pstate_1}, w_1 = \psems{\probzelusexpr_1}{\env_k}(\mem_1, \reveal{\pstate}, w)}\\ &&
  \letin{\env_1' = \env_k[y_1/v_1]}\\ &&
  \dots \\ &&
  \letin{v_{n}, \mem_{n}', \reveal{\pstate_n}, w_{n} = \psems{\probzelusexpr_n}{\env_{n-1}'}(\mem_{n}, \reveal{\pstate_{n-1}}, w_{n-1})}\\ &&
  \letin{\env_n' = \env_{n-1}'[y_{n}/v_{n}]}\\ &&
  \letin{v, \mem', \reveal{\pstate'}, w' = \psems{\probzelusexpr}{\env_n'}(\mem, \reveal{\pstate_n}, w_n)}\\ &&
  (v, ((\env_n'(x_1), \dots, \env_n'(x_k)), (\mem_1', \dots, \mem_n'), \mem'), \reveal{\pstate'}, w')
\end{array}$}
\caption{Delayed sampling semantics of ProbZelus programs. These semantics make use of the \textsc{assume}, \textsc{value}, and \textsc{observe} functions, which are the common interface for delayed sampling and semi-symbolic inference.
Pairs are automatically lifted to n-ary arguments in operator application.
Differences with Figure~\ref{fig:pf_sem} are highlighted.}
\label{fig:semisymb_sem}
\end{figure}

As an alternative to fully approximate particle filtering, ProbZelus programs may execute using \emph{delayed sampling}~\citep{murray_ds}, a technique for incorporating exact inference into the runtime inference system.
With delayed sampling, the key idea is as follows: instead of eagerly drawing samples at each invocation of the \zl{sample} operation, the system will execute lazily, delaying the sampling operation in the hope that it can find and exploit an opportunity to apply a known closed-form solution to the inference problem.
A complete exposition of delayed sampling in ProbZelus may be found in~\citet{rppl}.

Following~\citet{murray_ds}, we define the semantics in terms of a \emph{symbolic interface}, a set of functions that can be implemented to provide either delayed sampling, as presented in \citet{murray_ds,rppl}, or semi-symbolic inference, as presented in Section~\ref{sec:semisymb}.

\begin{definition}[Symbolic Interface]
The \emph{symbolic interface} consists of three functions that manipulate a symbolic state $g$.
The symbolic state $\pstate$ is a data structure that only interacts with the semantics through these functions.
\begin{itemize}
\item $\randomvar, \pstate' = \textsc{assume}(\mu, \pstate)$ returns a pair of a new random variable $\randomvar$ and a new symbolic state $\pstate'$ that has $\randomvar$ bound to a symbolic representation of the input distribution $\mu$.
\item $v, \pstate' = \textsc{value}(\randomvar, \pstate)$ samples the input random variable and returns the sampled value and the new symbolic state.
\item $\pstate', s = \textsc{observe}(\randomvar, v, \pstate)$ conditions the model on the fact that the input random variable $\randomvar$ takes on the value $v$, and returns the new state and the \emph{score}~(i.e., the value of the probability density) of $v$ under the marginal distribution of $\randomvar$.
\end{itemize}
\label{def:interface}
\end{definition}

For the purpose of this section, the symbolic state $\pstate$ is abstract. 
Different implementations provide either delayed sampling (e.g.\ \citet{murray_ds}) or semi-symbolic inference (Section~\ref{sec:semisymb}).

\paragraph{Semantics}

In Figure~\ref{fig:semisymb_sem}, we redefine the semantics of Figure~\ref{fig:pf_sem} to use delayed sampling constructs, by means of the symbolic interface.
The semantics of an expression $\psems{\probzelusexpr}{\env}$ is now a function that takes in a memory~$\mem$, a symbolic state $\pstate$, and a weight~$w$, and returns a symbolic value, an updated memory, a new symbolic state, and an updated weight.
The definition is similar to that in Figure~\ref{fig:pf_sem} except that
1) the symbolic state $\pstate$ is threaded throughout the semantics,
2) an operator application yields a symbolic term $\texttt{app}(\mathit{op},v)$ for the application of $\mathit{op}$ to the value $v$,
and 3) the semantics of \zl*sample*, \zl*observe*, and \zl*value* make use of the symbolic interface functions.

\paragraph{Inference.} Inference under delayed sampling~(DS) is also defined as a transformation on ProbZelus expressions.
It is defined as follows:
\[
\begin{array}{r@{~}c@{~}l}
\semi{\zl*infer*_{\mathit{DS}}(N, e)}\phantom{\,(\mem)} & = &
  \delta_{(\psemi{e}, \pstateempty)}
\\[0.5em]
\sems{\zl*infer*_{\mathit{DS}}(N, e)}{\env}(\mem) & = &
  \begin{array}[t]{@{}l@{}}
    \mathit{let}\; {
       \left[
       \begin{array}{@{}l@{}}
        d_i, \mem_i', \pstate_i', w_i' =
        \begin{array}[t]{@{}l@{}}
        \letin{\mem_i, \pstate_i = \textsc{draw}(\mem)} \\
        \letin{e_i, \mem_i', \pstate_i', w_i' =\psems{e}{\env}(\mem_i, \pstate_i, 1)} \\
        \textsc{distribution}(e_i, \pstate_i'), \mem_i', \pstate_i', w_i'
        \end{array}
        \end{array}
    \right]_{1 \le i \le N}}\\
    \letin{\mu = \lambda U.\ \Sigma_{i=1}^N \overline{w_i} * d_i(\pi_1(U))* \delta_{(\mem_i',\pstate_i')}(\pi_2(U))}\\
    (\pi_{1*}(\mu), \pi_{2*}(\mu))
  \end{array}
\end{array}
\]
The definition is similar to the particle-filter definition of \zl*infer*, but additionally threads the symbolic state $\pstate$ through the execution.
The initial memory is a Delta distribution of the initial memory of~$\probzelusexpr$ and the empty symbolic state~$\pstateempty$, and the output is a pair of a distribution of outputs and a distribution of memories and symbolic states.
The step function of each particle can return a symbolic expression containing un-sampled variables, and the \textsc{distribution} function returns the distribution corresponding to the symbolic expression $e_i$ without altering the symbolic state.

\paragraph{Correctness.} The operations in the symbolic interface must be designed such that the symbolic state represents the same distribution as the particle filter.
\citet{lunden17} proposed a sufficient condition for the delayed sampling symbolic state $\pstate$ to be correct.
Here we summarize the general idea behind this correctness condition and restate it for the symbolic interface in general. We leave the full formalization of correctness of the symbolic interface to Appendix~\ref{sec:proof_app} and in particular Theorem~\ref{thm:correctness}.

Let $\Pr(\Randomvar)$ be the distribution induced by \textsc{assume} statements on the set of random variables $\Randomvar$.
From this, for each random variable $\randomvar$, there is a conditional distribution $\Pr(\randomvar \mid \hat{V} = \hat{v}, \hat{O} = \hat{o})$ that specifies the \emph{marginal distribution} of $\randomvar$ conditioned on the random variables in $\hat{V}$ and $\hat{O}$ taking on the values $\hat{v}$ and $\hat{o}$, respectively.
We assume that $\hat{V}$ is the set of random variables that have previously been passed to $\textsc{value}$ and $\hat{O}$ is the set of random variables that have been passed to $\textsc{observe}$.
According to \citet{lunden17}, the symbolic interface functions are correct if:
\begin{itemize}
\item $\textsc{value}(\randomvar, \pstate)$ draws a sample from $\Pr(\randomvar \mid \hat{V} = \hat{v}, \hat{O} = \hat{o})$, and
\item $\textsc{observe}(\randomvar, r, \pstate)$ evaluates the probability density of $\Pr(\randomvar \mid \hat{V} = \hat{v}, \hat{O} = \hat{o})$ at the value $r$.
\end{itemize}
We provide a more detailed formalism that precisely specifies $\Pr(\randomvar \mid \hat{V} = \hat{v}, \hat{O} = \hat{o})$ in Appendix~\ref{sec:proof_app}.

\renewcommand{\plus}[2]{{\texttt{app(+,}{#1}\texttt{,}{#2}\texttt{)}}}
\renewcommand{\plusbold}[2]{{\texttt{app(+,}{#1}\texttt{,}{#2}\texttt{)}}}
\renewcommand{\minus}[2]{{\texttt{app(-,}{#1}\texttt{,}{#2}\texttt{)}}}
\renewcommand{\mult}[2]{{\texttt{app(*,}{#1}\texttt{,}{#2}\texttt{)}}}
\renewcommand{\division}[2]{{\texttt{app(/,}{#1}\texttt{,}{#2}\texttt{)}}}
\renewcommand{\squareroot}[1]{{\texttt{app(/,}{#1}\texttt{)}}}
\renewcommand{\ite}[3]{{\texttt{app(ite,}{#1}\texttt{,}{#2}\texttt{,}{#3}\texttt{)}}}

\section{Semi-Symbolic Inference}
\label{sec:semisymb}

\begin{figure}
\begin{align*}
\Distr \bnfarrow \; & \normal{\Expr}{\Expr}\;
\mid \; \bern{\Expr}\;
\mid \; \betad{\Expr}{\Expr} \;
\mid \; \deltad{\Expr} \\\\
\Expr \bnfarrow \; & \texttt{app(}\mathit{Op}\texttt{,}E^*\texttt{)}
\mid \; \real\;
\mid \; \integer\;
\mid \;  \randomvar\\\\
\mathit{Op} \bnfarrow \; & \texttt{+} \;
\mid \; \texttt{-} \;
\mid \; \texttt{*} \;
\mid \; \texttt{/} \;
\mid \; \texttt{sqrt} \;
\mid \; \texttt{ite} \;
\mid \; \texttt{=} \;
\mid \; \texttt{!=} \;
\mid \; \texttt{<}  \;
\mid \; \texttt{<=} \\\\
\real \in \Real,\ &\integer \in \Integer, \randomvar \in \Randomvar
\end{align*}
\caption{Grammar of symbolic expressions.}
\label{fig:symb-grammar}
\end{figure}

In this section, we present semi-symbolic inference as a series of definitions that implement the symbolic interface from Definition~\ref{def:interface}.
The section begins by defining the syntax of symbolic expressions and giving a definition of the symbolic state. It then implements the functions of the symbolic interface through a series of definitions:

\begin{itemize}
\item Section~\ref{sec:swap} defines the \textsc{swap} operation, the core building block of semi-symbolic inference that changes the dependency order between two random variables in the symbolic state.
\item Section~\ref{sec:partial-evaluator} presents the \textsc{eval} and \textsc{intervene} helper functions, which evaluate symbolic expressions and perform interventions on the symbolic state, respectively.
\item Section~\ref{sec:hoist} presents the \textsc{hoist} operation, which combines a sequence of swap operations to turn a given random variable into a root.
\item Section~\ref{sec:interface-impl} shows how to implement the interface in Definition~\ref{def:interface}.
\item Section~\ref{sec:semisymb-correctness} shows that these definitions satisfy correctness properties.
\end{itemize}

\paragraph{Symbolic Expressions}

Figure~\ref{fig:symb-grammar} gives the grammar of symbolic expressions used by semi-symbolic inference.
The grammar defines a set of symbolic distributions $\Distr$, consisting of a Gaussian distribution (denoted~$\mathcal{N}$), a Bernoulli distribution, a Beta distribution, and a Delta distribution.
The parameters for each distribution type are elements of a grammar of expressions $\Expr$.
An expression is either a real number $\real$, an integer $\integer$, a random variable $\randomvar$, or a n-ary operator $\mathit{Op}$ applied recursively to sub-expressions: $\texttt{app(}\textit{Op}\texttt{,}E^*\texttt{)}$.
The operators consist of standard arithmetic operators ($\texttt{+}$, $\texttt{-}$, $\texttt{*}$, and $\texttt{/}$), the square root operator \texttt{sqrt}, a conditional operator \texttt{ite}, and standard comparison operators ($\texttt{=}$, $\texttt{!=}$, $\texttt{<}$, and $\texttt{<=}$).
The interpretation of the conditional operator is that $\texttt{app(ite,}e_i\texttt{,}e_t\texttt{,}e_e\texttt{)}$ returns the value of the expression~$e_t$ if the condition expression $e_i$ evaluates to true, and otherwise returns the value of the expression $e_e$.

For example, $\normal{\randomvar_v - 2 * \randomvar_o}{1}$, the distribution of $\randomvar_l$ in Figure~\ref{fig:obsdiag1}, would be encoded in the grammar of Figure~\ref{fig:symb-grammar} as
$$\normal{\minus{\randomvar_v}{\mult{2}{\randomvar_o}}}{1}$$

\begin{definition}[Symbolic State] We define the \emph{symbolic state} of the semi-symbolic runtime inference system as a finite mapping from random variables to distributions $\pstate \in \Pstate = \Randomvar \mapsto D$.
\end{definition}

For example, the symbolic state depicted in Figure~\ref{fig:obsdiag1} is the finite map $$\pstate_{\mathit{example}} = \{\randomvar_v \mapsto \normal{0}{2500};\; \randomvar_o \mapsto \normal{0}{2500};\; \randomvar_l \mapsto \normal{\minus{\randomvar_v}{\mult{2}{\randomvar_o}}}{1}\}$$

\subsection{Swapping Random Variables}
\label{sec:swap}

\newcommand{\var}{\textit{var}}

\algnewcommand\algorithmicswitch{\textbf{switch}}
\algnewcommand\algorithmiccase{\textbf{case}}
\algnewcommand\algorithmicassert{\texttt{assert}}
\algnewcommand\Assert[1]{\State \algorithmicassert(#1)}%
\algdef{SE}[SWITCH]{Switch}{EndSwitch}[1]{\algorithmicswitch\ #1\ \algorithmicdo}{\algorithmicend\ \algorithmicswitch}%
\algdef{SE}[CASE]{Case}{EndCase}[1]{\algorithmiccase\ #1}{\algorithmicend\ \algorithmiccase}%
\algdef{SE}[OTHERWISE]{Otherwise}{EndOtherwise}{\algorithmicelse}{\algorithmicend\ \algorithmicelse}%
\algtext*{EndSwitch}%
\algtext*{EndCase}%
\algtext*{EndOtherwise}%

\begin{algorithm}
\renewcommand{\plus}[2]{{{#1} \; \texttt{+} \; {#2}}}
\renewcommand{\plusbold}[2]{{{#1} \oplus {#2}}}
\renewcommand{\minus}[2]{{{#1} \; \texttt{-} \; {#2}}}
\renewcommand{\mult}[2]{{{#1} \; \texttt{*} \; {#2}}}
\renewcommand{\division}[2]{\frac{{#1}}{{#2}}}
\renewcommand{\squareroot}[1]{{\sqrt{{#1}}}}
\renewcommand{\summation}[3]{{\texttt{sum} \; {#1} \; \texttt{in} \; {#2} \; \texttt{:} \; {#3}}}
\renewcommand{\ite}[3]{{\texttt{ite(} \; {#1}, {#2}, {#3} \; \texttt{)}}}

    \begin{algorithmic}
    \Function{swap}{$\randomvar_1$, $\randomvar_2$, $\pstate$}
    \Switch{$\pstate(\randomvar_1)$, $\pstate(\randomvar_2)$}
    \Case $\normal{\mu_0}{\var_0}, \normal{\mu}{\var}$ \textbf{if} $ \textsc{affine}(\mu, \randomvar_1) = (a, b) \land \textsc{const}(\var_0, \var)$
    \State $\mu_0', \var_0' \gets \plus{(\mult{a}{\mu_0})}{b}, \mult{(\mult{a}{a})}{\var_0}$
    \State $\var_0'', \mu_0'' \gets \division{1}{\plus{\division{1}{\var_0}}{\division{1}{\var}}}, \mult{\left(\plus{\division{\mu_0'}{\var_0'}}{\division{\randomvar_2}{\var}}\right)}{\var_0''}$
    \State \Return $(\pstate \begin{array}[t]{@{}l@{}}
        [\randomvar_1 \mapsto \normalbig{\division{\minus{\mu_0''}{b}}{a}}{\division{\var_0''}{\mult{a}{a}}}] \\[0em]
        [\randomvar_2 \mapsto \normalbig{\mu_0'}{\plus{\var_0'}{\var}}], \mathbf{true})
    \end{array}$
    \EndCase
    \Case $\betad{a}{b}$, $\bern{\randomvar_1}$
    \State \Return $(\pstate \begin{array}[t]{@{}l@{}}
        [\randomvar_1 \mapsto \betad{\plus{a}{(\ite{\randomvar_2}{1}{0})}}{\plus{b}{(\ite{\randomvar_2}{0}{1})}}]\\[0em]
        [\randomvar_2 \mapsto \bernbig{\division{a}{\plus{a}{b}}}], \mathbf{true})
    \end{array}$
    \EndCase
    \Case $\bern{p_1}$, $\bern{p_2}$
    \State $p_1' \gets \division{\mult{p_1}{\ite{\randomvar_2}{p_2}{(\minus{1}{p_2})}[\randomvar_1 \leftarrow 1]}}{\ite{\randomvar_2}{p_2'}{(\minus{1}{p_2'})}}$
    \State $p_2' \gets \plus{(\mult{p_1}{p_2[\randomvar_1 \leftarrow 1]})}{(\mult{(\minus{1}{p_1})}{p_2[\randomvar_1 \leftarrow 0]})}$
    \State \Return $(\pstate \begin{array}[t]{@{}l@{}}
        [\randomvar_1 \mapsto \bern{p_1'}]\\[0em]
        [\randomvar_2 \mapsto \bern{p_2'}], \mathbf{true})
    \end{array}$
    \EndCase
    \Otherwise
    \State \Return $(\pstate, \mathbf{false})$
    \EndOtherwise
    \EndSwitch
    \EndFunction
    \end{algorithmic}
\renewcommand{\plus}[2]{{\texttt{app(+,}{#1}\texttt{,}{#2}\texttt{)}}}
\renewcommand{\plusbold}[2]{{\texttt{app(+,}{#1}\texttt{,}{#2}\texttt{)}}}
\renewcommand{\minus}[2]{{\texttt{app(-,}{#1}\texttt{,}{#2}\texttt{)}}}
\renewcommand{\mult}[2]{{\texttt{app(*,}{#1}\texttt{,}{#2}\texttt{)}}}
\renewcommand{\division}[2]{{\texttt{app(/,}{#1}\texttt{,}{#2}\texttt{)}}}
\renewcommand{\squareroot}[1]{{\texttt{app(/,}{#1}\texttt{)}}}
\renewcommand{\ite}[3]{{\texttt{app(ite,}{#1}\texttt{,}{#2}\texttt{,}{#3}\texttt{)}}}

    \caption{The definition of \textsc{swap}, incorporating exact inference for Gaussian distributions, Beta distributions, and Bernoulli distributions. Note that this algorithm uses shorthand notation for symbolic expression construction, e.g. $\expr_1 + \expr_2$ constructs a symbolic addition term $\plus{\expr_1}{\expr_2}$.}
    \label{alg:swap}
\end{algorithm}

The core operation of semi-symbolic inference is the \emph{swap}.
Swapping random variables changes the probabilistic dependence structure of the symbolic state without changing the overall distribution the symbolic state represents.

\paragraph{Requirements}

To swap two random variables $\randomvar_1$ and $\randomvar_2$, the following must hold:
\begin{enumerate}
  \item $\randomvar_1$ is a parent of $\randomvar_2$ in the initial state, making $\randomvar_2$ a parent of $\randomvar_1$ in the new state.
  \item After the swap, all variables other than $\randomvar_1$ and $\randomvar_2$ have the same distributions as before.
  \item The symbolic state represents the same overall joint distribution before and after the swap.
\end{enumerate}

\paragraph{Dependency Cycles}
Furthermore, a swap is only legal in some circumstances because a swap may introduce new probabilistic dependencies between random variables.
In general, a swap will introduce dependencies between each of $\randomvar_1$ and $\randomvar_2$ and all parents of either $\randomvar_1$ or $\randomvar_2$.
A legal swap is one that does not create dependency cycles between random variables.
The function $\textsc{can\_swap}(\randomvar_1, \randomvar_2, \pstate)$ determines if the swap of $\randomvar_1$ and $\randomvar_2$ is legal in symbolic state $\pstate$, i.e.\ whether or not swapping $\randomvar_1$ and $\randomvar_2$ will introduce a dependency cycle.

\paragraph{Swap Algorithm}

Algorithm~\ref{alg:swap} defines the \textsc{swap} function for Gaussians, Beta distributions, and Bernoulli distributions.
The function takes as input a parent random variable $\randomvar_1$, a child random variable $\randomvar_2$, and an initial state $\pstate$.
It returns a modified state and a boolean indicating whether or not a swap is possible using closed-form solutions known to the semi-symbolic inference system.

Note that there is a difference between a swap being \emph{illegal} due to dependency cycles and being \emph{impossible} due to the semi-symbolic inference system not identifying a closed-form solution. The function \textsc{swap} returns \textbf{false} if no closed-form solution exists and a swap is impossible, whereas \textsc{can\_swap} returns \textbf{false} if the swap is illegal due to dependency cycles.

Also note that for ease of presentation, Algorithm~\ref{alg:swap} uses shorthand notation for symbolic expression construction.
In particular, Algorithm~\ref{alg:swap} uses $\expr_1 + \expr_2$, $\expr_1 - \expr_2$, $\expr_1 * \expr_2$, $\frac{\expr_1}{\expr_2}$, and $\texttt{ite}(\expr_i, \expr_t, \expr_e)$ as shorthand for $\plus{\expr_1}{\expr_2}$, $\minus{\expr_1}{\expr_2}$, $\mult{\expr_1}{\expr_2}$, $\division{\expr_1}{\expr_2}$, and $\ite{\expr_i}{\expr_t}{\expr_e}$, respectively.
We stress that Algorithm~\ref{alg:swap} performs no numerical computation, and that these operations are constructors for symbolic expressions.

The implementation of \textsc{swap} is separated into three cases:

\paragraph{Gaussian}

The first case occurs when both $\randomvar_1$ and $\randomvar_2$ are Gaussian-distributed, and also requires the variance of each distribution to be constant (i.e., to not depend on any random variables) and for the mean of $\randomvar_2$ to be expressible as an affine function of $\randomvar_1$.
We express the analysis for this condition as $\textsc{affine}(\mu, \randomvar_1) = (a, b)$, which means that $\mu$ can be written as $a * \randomvar_1 + b$ where $a$ and $b$ are themselves symbolic expressions that may contain random variables other than $\randomvar_1$.

This case applies the following rules for Gaussian distributions: 1) the symbolic expressions $\mu_0'$ and $\var_0'$ represent a linear transformation of the Gaussian distribution of $\randomvar_1$, which is also a Gaussian distribution; 2) the expressions $\mu_0''$, $\var_0''$ are computed from the rules for conjugate priors~\citep{conjugate_priors}; 3) the final expressions for the new distributions of $\randomvar_1$ and $\randomvar_2$ incorporate both the inverse linear transformation to the one used to generate $\mu_0'$ and $\var_0'$.

Using these rules, the symbolic state is updated and returned.
The notation $\pstate[\randomvar \mapsto \distr]$ means that the symbolic state $\pstate$ is updated with the random variable $\randomvar$ remapped to the distribution $\distr$.

\paragraph{Example} The example in Section~\ref{sec:example} makes extensive use of swaps between Gaussian distributions.
Here, we briefly discuss how \textsc{swap} executes to produce the distribution for $\randomvar_l$ in Figure~\ref{fig:obsdiag2}.
The runtime inference system produces the symbolic state depicted in Figure~\ref{fig:obsdiag2} from one depicted in Figure~\ref{fig:obsdiag1} by swapping the random variable $\randomvar_v$ with its child $\randomvar_l$.
Formally, it executes $\textsc{swap}(\randomvar_v, \randomvar_l, \pstate_{\mathit{example}})$, where $\pstate_\mathit{example}$ is the symbolic state depicted in Figure~\ref{fig:obsdiag1}.

This falls into the Gaussian case, with $\mu_0 = 0$, $\var_0 = 2500$, $\mu = \minus{\randomvar_v}{\mult{2}{\randomvar_o}}$, and $\var = 1$.
The affine analysis concludes that the expression for $\mu$ can be written as $\mu = 1 * \randomvar_v + (-2 * \randomvar_o)$, and thus produces $a = 1$ and $b = \mult{-2}{\randomvar_o}$.
Next, the \textsc{swap} function uses these values of $\mu_0$, $\var_0$, $a$, and $b$ to compute that $\mu_0' = \plus{\mult{1}{0}}{\mult{-2}{\randomvar_o}}$ and $\var_0' = \mult{\mult{1}{1}}{2500}$.
Ultimately, using these values of $\mu_0'$, $\var_0'$, and $\var$, the \textsc{swap} function will return a symbolic state in which $\randomvar_l$ maps to the symbolic expression
$$\normal{\plus{\mult{1}{0}}{\mult{-2}{\randomvar_o}}}{\plus{\mult{\mult{1}{1}}{2500}}{1}}$$
Similarly, the \textsc{swap} function will compute the symbolic expressions $\mu_0''$ and $\var_0''$ and use them to produce a new distribution for $\randomvar_v$.

\paragraph{Beta-Bernoulli}

The second case occurs when $\randomvar_1$ is distributed according to a Beta distribution and $\randomvar_2$ is distributed according to a Bernoulli distribution.
This is an instance of the Beta-Bernoulli conjugate model, with its own rules and transformations~\citep{conjugate_priors}.

\paragraph{Bernoulli-Bernoulli}

The third case applies when both $\randomvar_1$ and $\randomvar_2$ have Bernoulli distributions.
This case applies the rules of discrete probability to produce a new symbolic state.
In particular, it sums out the random variable $\randomvar_1$ to produce an expression for the new probability $p_2'$ of $\randomvar_2$.
It then uses Bayes' rule to produce a new expression for $\randomvar_1$.
Note that this definition uses the notation $\expr_1[\randomvar \leftarrow \expr_2]$ to denote expression $\expr_1$ with expression $\expr_2$ substituted for the random variable $\randomvar$.

\paragraph{Extensibility}

In Algorithm~\ref{alg:swap}, if none of these cases apply, \textsc{swap} returns the existing state and the value false.
However, the definition of \textsc{swap} is extensible and can handle additional cases as necessary.
In our implementation, we have added cases for multivariate Gaussians and categorical distributions.
Additional cases for more conjugate priors (e.g.\ those in \citet{conjugate_priors}) may follow the example of the Beta-Bernoulli conjugacy from the second case of Algorithm~\ref{alg:swap}.

\subsection{Evaluation and Intervention} \label{sec:partial-evaluator}

\paragraph{Evaluation}

To simplify symbolic expressions introduced by operator applications, semi-symbolic inference makes use of a \emph{partial evaluator}: a function $\textsc{eval}(\expr, \pstate) \in \Expr \times \Pstate \mapsto \Expr$ that takes in an expression and a symbolic state and produces a new expression that is evaluated to a constant if possible.
The partial evaluator is defined recursively on the structure of expressions.
For example, on addition, the evaluator proceeds according to:
$$
\textsc{eval}(\plus{\expr_1}{\expr_2}, \pstate) =
\begin{cases}
\real_1 + \real_2 & \textsc{eval}(\expr_1, \pstate) = \real_1 \land \textsc{eval}(\expr_2, \pstate) = \real_2\\
\plus{\textsc{eval}(\expr_1, \pstate)}{\textsc{eval}(\expr_2, \pstate)} & \text{otherwise}
\end{cases}
$$
This equation specifies that if both subexpressions $\expr_1$ and $\expr_2$ evaluate to real numbers, the partial evaluator performs real-number arithmetic and returns the result.
Otherwise, it recursively partially evaluates the subexpressions and leaves the result symbolic.

For example, the variance of the distribution for $\randomvar_l$ in Figure~\ref{fig:obsdiag2} is the symbolic expression $\plus{2500}{1}$.
Letting $\pstate'_\mathit{example}$ be a depiction of the symbolic state in Figure~\ref{fig:obsdiag2}, this expression evaluates to $\textsc{eval}(\plus{2500}{1}, \pstate'_\mathit{example}) = 2501$.
By contrast, the mean of this distribution is $\minus{0}{\mult{2}{X_o}}$, which cannot be evaluated any further, which means that $\textsc{eval}(\minus{0}{\mult{2}{X_o}}, \pstate'_\mathit{example}) = \minus{0}{\mult{2}{X_o}}$.

One unique feature of the semi-symbolic partial evaluator is how it handles Delta distributions.
In particular, the evaluator leverages the fact that Delta-distributed variables must only take on one value and therefore can be substituted like normal program variables.
The partial evaluator handles Deltas according to the equation:
$$
\textsc{eval}(\randomvar, \pstate) = \begin{cases}
\textsc{eval}(e, \pstate) & \pstate(\randomvar) = \deltad{\expr} \\
\randomvar & \text{otherwise}
\end{cases}
$$
This equation states that if a random variable has a Delta distribution, then the partial evaluator evaluates the Delta's internal expression and returns the result.
For example, in Figure~\ref{fig:intervenediag}, the distribution for $\randomvar_o$ is only a function of constants and $\randomvar_l$, and $\randomvar_l$ has a Delta distribution.
Formally, if $\pstate'''_\mathit{example}$ is the symbolic state depicted in Figure~\ref{fig:intervenediag}, then $\textsc{eval}(\pstate'''_\mathit{example}(\randomvar_o), \pstate'''_\mathit{example}) = \normal{0.4}{500}$.

We also define a version of the partial evaluator that updates the symbolic state, which we write $\textsc{eval}^*(\randomvar, \pstate) \in \Randomvar \times \Pstate \mapsto \Pstate$.
This evaluator updates each parameter of a given random variable's distribution to be its partially evaluated counterpart.
For example, for Gaussian distributions:
$$
\textsc{eval}^*(\randomvar, \pstate) = \pstate[\randomvar \mapsto \normal{\textsc{eval}(\mu, \pstate)}{\textsc{eval}(\var, \pstate)}] \; \mathrm{if} \; \pstate(\randomvar) = \normal{\mu}{\var}
$$

\paragraph{Intervention}

An \emph{intervention} replaces a root node with a Delta distribution.
Intervention is defined as the function $\textsc{intervene} \in \Randomvar \times \Val \times \Pstate \mapsto \Pstate$ that takes in a random variable, the intervention value, the symbolic state, and returns the new symbolic state.
$\Val$ denotes the sample space of the random variable. 
The function is defined as follows:
$$
\textsc{intervene}(\randomvar, \real, \pstate) = \pstate[\randomvar \mapsto \deltad{\real}]
$$
In Figure~\ref{fig:symb_diag} the transition from \ref{fig:obsdiag3} to \ref{fig:intervenediag} depicts the execution of $\pstate'''_\mathit{example} = \textsc{intervene}(\randomvar_l, -1, \pstate''_\mathit{example})$, where $\pstate''_\mathit{example}$ is the symbolic state depicted in Figure~\ref{fig:obsdiag3} and $\pstate'''_\mathit{example}$ is the one depicted in Figure~\ref{fig:intervenediag}.

\subsection{Hoisting}
\label{sec:hoist}

Swaps are composed together to support the implementation of the operations from Definition~\ref{def:interface} by an operation called \textsc{hoist}.
The objective of hoisting a random variable is to update the symbolic state so that the variable in question is a \emph{root} that depends on no other random variables.

\paragraph{Preliminaries}

The \textsc{hoist} operation depends on functions manipulating lists of random variables:
\begin{itemize}
    \item $\textsc{get\_parents}(\randomvar, \pstate)$ returns a list of parents of $\randomvar$ in $\pstate$, i.e. a list of random variables that are free variables in the expression of the distribution of $\randomvar$.
    \item $\textsc{topo\_sort}(\hat{\randomvar}, \pstate)$ sorts a list of random variables in topological order according to the parent-child relation of $\pstate$.
    \item $\textsc{reverse}(\hat{\randomvar})$ reverses a list of random variables.
\end{itemize}

\begin{algorithm}
\begin{algorithmic}
\Function{hoist\_helper}{$\randomvar_\mathrm{cur}$, \texttt{roots}, $\pstate$}
\State \texttt{parents} $\gets$ $\textsc{topo\_sort}(\textsc{get\_parents}(\randomvar_\mathrm{cur}, \pstate))$
\State \texttt{roots}' $\gets$ \texttt{roots};\; $\pstate' \gets \pstate$
\For{$\randomvar_\mathrm{par} \in $ \texttt{parents}}
\If{$\randomvar_\mathrm{par} \not \in $ \texttt{roots}}
\State $\pstate' \gets$ \Call{hoist\_helper}{$\randomvar_\mathrm{par}$, \texttt{roots}', $\pstate'$}$;\;$ \texttt{roots}' $\gets \randomvar_\mathrm{par} ::$ \texttt{roots}'
\EndIf
\EndFor
\State $\pstate'' \gets \pstate'$
\For{$\randomvar_\mathrm{par} \in $ \textsc{reverse}(\texttt{parents})}
\If{$\randomvar_\mathrm{par} \not \in $ \texttt{roots}}
\State \textbf{assert} \Call{can\_swap}{$\randomvar_\mathrm{par}$, $\randomvar_\mathrm{cur}$, $\pstate''$}$;\;$ $(\pstate'', \texttt{conjugate}) \gets$ \Call{swap}{$\randomvar_\mathrm{par}$, $\randomvar_\mathrm{cur}$, $\pstate''$}
\If{not \texttt{conjugate}}
\State \textbf{throw} $(\randomvar_\mathrm{par}, \randomvar_\mathrm{cur})$
\EndIf
\EndIf
\EndFor
\State \Return $\pstate''$
\EndFunction
\Function{hoist}{$\randomvar_\mathrm{in}$, $\pstate$}
\try
\State \Return \Call{hoist\_helper}{$\randomvar_\mathrm{in}$, $\{\}$, $\pstate$}
\catch{$(\randomvar_\mathrm{par}, \randomvar_\mathrm{child})$}
\State $(\_, \pstate') \gets$ \Call{\textsc{value}}{$\randomvar_{\mathrm{par}}, \pstate$}$;\;$ $\pstate'' \gets \textsc{eval}^*(\randomvar_\mathrm{child}, \pstate');\; $ \Return \Call{hoist}{$\randomvar_\mathrm{in}$, $\pstate''$}
\endtry
\EndFunction
\end{algorithmic}
\caption{Hoisting a random variable to be a root depending on no other random variables.}
\label{alg:hoist}
\end{algorithm}

\paragraph{Hoisting Algorithm}

Algorithm~\ref{alg:hoist} defines $\textsc{hoist} \in \Randomvar \times \Pstate \mapsto \Pstate$.
It makes use of a helper function called \textsc{hoist\_helper} that takes a set of root variables given by the parameter \texttt{roots}.
\textsc{hoist\_helper}'s objective is to turn the input variable $\randomvar_\mathrm{cur}$ into a root variable, except that variables that are in \texttt{roots} do not count for the purpose of determining whether or not $\randomvar_\mathrm{cur}$ is a root variable.

To do so, $\textsc{hoist\_helper}$ first recursively calls itself on all parents of $\randomvar_\mathrm{cur}$ in topological order.
The \textsc{topo\_sort} function yields an order such that the first element of the resulting \texttt{parents} list has no ancestors that are also in \texttt{parents}.
Then, because on subsequent recursive calls, all previously visited parents are added to \texttt{roots} and are thus excluded from being hoisted or swapped, later elements of \texttt{parents} will be descendants of earlier elements of \texttt{parents} after the recursive calls.

After the recursion, the function iterates through all parents in reverse topological order.
This reverse ordering ensures that the algorithm can always swap each parent with $\randomvar_\mathrm{cur}$ without creating a cycle.
This is because in order to create a cycle, the algorithm would need to swap the child node with a parent whose descendant is the child node itself.
By iterating in reverse topological order, any other parent that would enable such a path to exist in the dependency graph must already have been swapped and therefore does not have $\randomvar_\mathrm{child}$ as a dependency.
The \textsc{can\_swap} assertion encapsulates the above argument that the algorithm does not create cycles. In Appendix~\ref{sec:proof_side_conditions}, we formally prove that this assertion always passes at runtime.

In case the distributions are not conjugate and therefore a swap is impossible, as indicated by the \texttt{conjugate} variable being \textbf{false}, the algorithm throws an exception that is caught at the outermost level.
It then calls the \textsc{value} function (from Definition~\ref{def:interface} and defined below in Section~\ref{sec:interface-impl}) to replace the parent variable with a random sample.
It next evaluates the child random variable to eliminate the resulting Delta distribution (thus eliminating the need to perform this swap) and finally restarts the hoisting process from the beginning.

\paragraph{Example} In the example in Figure~\ref{fig:symb_diag}, Figures~\ref{fig:obsdiag1}--\ref{fig:obsdiag3} depict the result of executing $\textsc{hoist}(\randomvar_l, \pstate_\mathit{example})$ where $\pstate_\mathit{example}$ is the symbolic state depicted in Figure~\ref{fig:obsdiag1}.
The function $\textsc{hoist}$ immediately calls $\textsc{hoist\_helper}$ with $\randomvar_\mathrm{cur} = \randomvar_l$ and $\texttt{roots} = \{\}$.
The first stage of $\textsc{hoist\_helper}$ is to recursively call $\textsc{hoist\_helper}$ on its ancestors in topological order.
The ancestors of $\randomvar_l$ are $\randomvar_o$ and $\randomvar_v$, and as they have no dependencies between each other, any order is a valid topological order.
Furthermore, because neither $\randomvar_o$ nor $\randomvar_v$ has a parent, calling $\textsc{hoist\_helper}$ on these variables has no effect.
After all recursive calls, $\textsc{hoist\_helper}(\randomvar_l, \{\})$ swaps $\randomvar_l$ with both $\randomvar_o$ and $\randomvar_v$.
In the example in Figure~\ref{fig:symb_diag}, we assume that $\textsc{hoist\_helper}$ first swaps with $\randomvar_v$ (Figure~\ref{fig:obsdiag2}) and then with $\randomvar_o$ (Figure~\ref{fig:obsdiag3}).
After these swaps, $\textsc{hoist}$ returns the resulting symbolic state.

\subsection{Symbolic Interface}
\label{sec:interface-impl}

In this section, we describe how to implement the symbolic interface presented in Definiton~\ref{def:interface}.

\paragraph{Assume} The operation $\textsc{assume} \in \Distr \times \Pstate \mapsto \Randomvar \times \Pstate$ takes in a distribution and a current state, and returns a new random variable and the updated state.
It is defined as follows:
$$
\textsc{assume}(\distr, \pstate) = (\randomvar_\mathrm{new}, \pstate[\randomvar_\mathrm{new} \mapsto \distr]) \; \mathrm{where} \; \randomvar_\mathrm{new} \; \textnormal{is not assigned in} \; \pstate
$$
This definition specifies that the \textsc{assume} function returns a fresh random variable, and that it updates the symbolic state to have the new random variable point to the input distribution.

\paragraph{Value}
The function $\textsc{value} \in \Randomvar \times \Pstate \mapsto \Pstate \times \Val$ instructs the inference system to replace a particular random variable with a sample from its marginal distribution and return the resulting sample.
It is defined as follows, and is mutually recursive with the \textsc{hoist} operation:
\begin{align*}
\textsc{value}(\randomvar, \pstate) = \; & \mathit{let} \; \pstate' = \textsc{hoist}(\randomvar, \pstate) \; \mathit{in}\\
& \mathit{let} \; \pstate'' = \textsc{eval}^*(\randomvar, \pstate') \; \mathit{in}\\
& \mathit{let} \; v = \textsc{draw}(\pstate''(\randomvar)) \; \mathit{in}\\
& (v, \textsc{intervene}(\randomvar, v, \pstate''))
\end{align*}
This definition specifies that the \textsc{value} function first hoists the variable $\randomvar$, guaranteeing that in the resulting symbolic state it is a root.
After hoisting, it evaluates the distribution of the random variable that was just hoisted, producing a closed-form distribution that it then samples from.
It further updates the symbolic state by intervention.

\paragraph{Observe}
The function $\textsc{observe} \in \Randomvar \times \Val \times \Pstate \mapsto \Pstate \times \Real$ conditions the symbolic state on the input random variable taking on the input value and returns the updated state and a score which corresponds to how likely this value is according to its marginal density.
It is defined as follows:
\begin{align*}
\textsc{observe}(\randomvar, v, \pstate) = \; & \mathit{let} \; \pstate' = \textsc{hoist}(\randomvar, \pstate) \; \mathit{in}\\
& \mathit{let} \; \pstate'' = \textsc{eval}^*(\randomvar, \pstate') \; \mathit{in}\\
& \mathit{let} \; s = \textsc{score}(\pstate''(\randomvar), v) \; \mathit{in}\\
& \mathit{let} \; \pstate''' = \textsc{intervene}(\randomvar, v, \pstate'') \; \mathit{in} \\
& (\pstate''', s)
\end{align*}
This definition specifies that the \textsc{observe} operation first hoists the input random variable and fully evaluates its distribution's parameters as it is now a root.
It then calculates the probability density of the variable's marginal distribution using the \textsc{score} function.
It returns a combination of the new state, obtained by intervening to condition the symbolic state on the fact that the random variable takes on the specified value, and the new weight.

\subsection{Correctness}
\label{sec:semisymb-correctness}

In this section, we formalize the correctness of semi-symbolic inference.
We present the key ideas necessary to specify and prove the correctness of each operation defined in the previous sections, leaving the full formalization to Appendix~\ref{sec:proof_app}. 

\paragraph{Swap Correctness} We first present the correctness requirements for \textsc{swap}. The \textsc{swap} function is correct if it preserves the joint distribution of all random variables. 
This statement requires defining $\sem{\pstate}$, the joint distribution that is the meaning of the symbolic state $\pstate$ .
We defer this definition to Appendix~\ref{sec:proof_app}, and formalize the correctness property as follows:
\begin{lemma}[Swap Preservation]
If $(\pstate', \_) = \textsc{swap}(\randomvar_1, \randomvar_2, \pstate)$, then $\sem{\pstate'} = \sem{\pstate}$.
\label{lem:swap_preservation}
\end{lemma}
The proof of the theorem is also in Appendix~\ref{sec:proof_app}.
The key idea is to incorporate known results about conjugate priors~\citep{conjugate_priors}.

\paragraph{Evaluation Correctness} Similarly, the $\textsc{eval}^*$ operation must also preserve the joint distribution:
\begin{lemma}[$\textsc{eval}^*$ Correctness]
If $\pstate' = \textsc{eval}^*(\randomvar, \pstate)$, then $\sem{\pstate'} = \sem{\pstate}$.
\label{lem:eval_star}
\end{lemma}

\paragraph{Intervention Correctness} The key idea for \textsc{intervene} is that if the random variable passed to it is a root, then it should perform conditioning.
In this theorem, we use the notation $\Pr_\pstate$ to refer to probability distributions implied by the symbolic state~$\pstate$.
We construct conditional probability distributions from the overall joint distribution $\sem{\pstate}$ using standard techniques; see Appendix~\ref{sec:proof_app} for details.
We formalize this idea as follows:
\begin{lemma}[\textsc{intervene} Correctness]
If $\randomvar$ is a root in $\pstate$, and $\pstate' = \textsc{intervene}(\randomvar, r, \pstate)$, then for any subset $\mathcal{V}'$ of random variables mapped in $\pstate$, $\Pr_{\pstate'}(\mathcal{V}') = \Pr_{\pstate}(\mathcal{V}' \mid \randomvar = r)$.
\label{lem:intervention}
\end{lemma}

\paragraph{Hoist Correctness}
We next present the correctness of the subroutine \textsc{hoist\_helper}.
This subroutine must preserve the semantics of the overall joint distribution.
It is also designed to turn its input variable into a root, with the exception of variables in \texttt{roots}.
We formalize this as follows:
\begin{lemma}[\textsc{hoist\_helper} Correctness]
If $\pstate' = \textsc{hoist\_helper}(\randomvar_\mathrm{cur}, \texttt{roots}, \pstate)$, and no exceptions are thrown, $\sem{\pstate'} = \sem{\pstate}$, and in $\pstate'$, $\randomvar_\mathrm{cur}$ is a root, except it may depend on variables in \texttt{roots}.
\label{lem:hoist_helper}
\end{lemma}
Next, we establish the correctness of \textsc{hoist}.
Due to the fact that \textsc{hoist} and \textsc{value} are mutually recursive, we combine their correctness properties into a single theorem.
The correctness theorem for \textsc{hoist} states that hoist turns the input random variable into a root and preserves the symbolic state, except that the new symbolic state will be conditioned on any variables that needed to be sampled due to lack of conjugacy.
The correctness theorem for \textsc{value} states that it produces a random sample from the appropriate marginal distribution, and conditions the new symbolic state on this and any other sampled variables.
\begin{theorem}[\textsc{hoist} and \textsc{value} Correctness]
If $\pstate' = \textsc{hoist}(\randomvar_\mathrm{in}, \pstate)$, then $\Pr_{\pstate'}(\Randomvar) = \Pr_\pstate(\Randomvar \mid \hat{V} = \hat{v})$, where $\hat{V}$ is the set of variables sampled during the execution of \textsc{hoist} and $\hat{v}$ is the corresponding set of sampled values.
Furthermore, after executing \textsc{hoist}, $\randomvar_\mathrm{in}$ is a root in $\pstate'$.

Also, if $(\pstate', v) = \textsc{value}(\randomvar, \pstate)$, then $v$ is a sample from $\Pr_g(\randomvar \mid \hat{V} = \hat{v})$ and $\Pr_{\pstate'}(\Randomvar) = \Pr_g(\Randomvar \mid \randomvar = v, \hat{V} = \hat{v})$, where $\hat{V}$ and $\hat{v}$ are as above.
\label{thm:hoist_value}
\end{theorem}

\paragraph{Observe Correctness} The final operation whose correctness we must ensure is \textsc{observe}.
This operation is correct if it conditions the symbolic state on the input variable being equal to the input value.
It further must return the density of the random variable's marginal distribution evaluated at the input value.
We formalize this as follows:
\begin{theorem}[\textsc{observe} Correctness]
If $(\pstate', w) = \textsc{observe}(\randomvar, r, \pstate)$, then we have that $\Pr_{\pstate'}(\Randomvar) = \Pr_{\pstate}(\Randomvar \mid \randomvar = r, \hat{V} = \hat{v})$, where $\hat{V}$ and $\hat{v}$ are the random variables that may need to be sampled during the observation, and $\hat{v}$ are their sampled values.
Furthermore, $w$ is the density of $\Pr_{\pstate}(\randomvar \mid \hat{V} = \hat{v})$.
\label{thm:observe}
\end{theorem}

\paragraph{Overall Correctness} As we explain in Section~\ref{sec:semisymb_sec}, \citet{lunden17} provides conditions on the symbolic interface that are sufficient to ensure the overall correctness of the inference algorithms.
These conditions depend, in general, on the total sequence of interface operations performed.
We provide a detailed formalism of the overall correctness of the semi-symbolic operations of the symbolic interface in Appendix~\ref{sec:semisymb_correctness_proofs}.

\section{Closed-family Properties of Semi-symbolic Inference}
\label{sec:properties}

In this section, we discuss the properties of semi-symbolic inference on \emph{closed families}.
Closed families are sets of symbolic states on which the semi-symbolic operations from Section~\ref{sec:semisymb} are guaranteed to maintain a symbolic representation without falling back on sampling-based approximations.
Developers can use this guarantee to reliably write probabilistic programs that the semi-symbolic runtime inference system will automatically implement as Rao-Blackwellized particle filters.
In this section, we define closed families and formalize their properties.
The definition of a closed family that it should be closed under all legal swaps:
\begin{definition}[Closed Family]
A closed family $\mathcal{C}$ is a set of symbolic states such that if $\pstate \in \mathcal{C}$, and $\randomvar_1, \randomvar_2$ in $\pstate$ such that $\textsc{can\_swap}(\randomvar_1, \randomvar_2)$, then $\textsc{swap}(\randomvar_1, \randomvar_2, \pstate) = (\pstate', \textbf{true})$ and $\pstate' \in \mathcal{C}$.
\end{definition}
For example, all the states depicted in Figure~\ref{fig:symb_diag} are in the linear-Gaussian closed family we define below.
This means that any legal swap we may want to execute on one of these states will be possible (i.e.\ will find available conjugate distributions), and furthermore will not modify the symbolic state in such a way that future swaps could be impossible.
Moreover, this means that the hoisting we perform in Figures~\ref{fig:obsdiag1}--\ref{fig:obsdiag3} does not perform any random sampling.

We now explain two classes of distributions and show they are closed families: linear-Gaussian distributions and finite discrete distributions.
\begin{theorem}[Linear-Gaussian closed family]
The set of linear-Gaussian symbolic states, consisting of states such that
a) all distributions in the state are Gaussian with constant variance (i.e., the variance does not depend on any random variables), and
b) the mean of each Gaussian distribution is an affine function of other random variables,
is a closed family.
\label{thm:gauss_cf}
\end{theorem}
\begin{theorem}[Finite discrete closed family]
The set of finite discrete symbolic states, consisting of states such that every distribution is a Bernoulli random variable, is a closed family.
\end{theorem}

The key step to prove that the linear-Gaussian family is closed is showing that the new means for the distributions (i.e.\ $\frac{\mu_0'' - b}{a}$ and $\mu_0'$ in Algorithm~\ref{alg:swap}) are affine and can be analyzed as such by the $\textsc{affine}$ analysis.
This is a constraint of the \textsc{affine} analysis's precision.
In our implementation, we use a recursive analysis that meets this constraint, but we do not formalize it here.
The remainder of the proofs of these theorems follow straightforwardly from the definition of \textsc{swap}.

\paragraph{RBPF Guarantee} The goal of closed families is to enable developers to control when random sampling happens over the course of semi-symbolic inference.
In particular, developers want to ensure the runtime does not perform any hidden calls to \textsc{value} through the catch clause of \textsc{hoist}, which would occur when there is no conjugacy available to \textsc{swap}.
To ensure this condition, we present the following Rao-Blackwellized particle filtering (RBPF) guarantee of the semi-symbolic implementation of the symbolic interface.
This theorem states conditions under which variables \textsc{assume}d from within the closed family are guaranteed to be exact.
In particular, all variables must either be in the closed family, or immediately sampled by being passed to \textsc{value}.

\begin{theorem}[RBPF Guarantee]
Given a closed family $\mathcal{C}$, let $\hat{\randomvar}_e$ be the set of random variables the developer wants to keep exact, and $\hat{\randomvar}_s = \Randomvar \setminus \hat{\randomvar}_e$ the variables the developer wants to sample.
If,
\begin{itemize}

\item For all calls $\randomvar, \pstate' = \textsc{assume}(\mu, \pstate)$ such that $\randomvar \in \hat{\randomvar}_e$, if $\pstate \in \mathcal{C}$ then $\pstate' \in \mathcal{C}$.
\item For all calls $\randomvar, \pstate' = \textsc{assume}(\mu, \pstate)$ such that $\randomvar \in \hat{\randomvar}_s$, $\randomvar$ is immediately passed to \textsc{value}, and $\mu$ does not depend on any variables in $\hat{\randomvar}_e$, then
\end{itemize}
all variables in $\hat{\randomvar}_s$ will be sampled by \textsc{value}, and no variable in $\hat{\randomvar}_e$ will ever be passed to \textsc{value}.
\label{thm:rbpf}
\end{theorem}

\paragraph{Proof Sketch} The key step of the proof is to show that the symbolic state will be in $\mathcal{C}$ during the execution of \textsc{hoist}.
Then, by the definition of closed families, \textsc{swap} will always return true and \textsc{hoist\_helper} will never throw an error.

\section{Evaluation}
\label{sec:evaluation}

In this section, we evaluate our implementation of semi-symbolic inference~(\ssdsacro) in the ProbZelus streaming probabilistic programming language.
We compare this new algorithm with the two main inference algorithms previously implemented in ProbZelus~\citep{rppl}, particle filtering~(PF) and delayed sampling~(\gcdsacro).
We address the following research questions:
\begin{description}[left=0pt]
\item[RQ1] Does the new algorithm provide more accurate results?
\item[RQ2] How fast is the new algorithm?
\end{description}

\subsection{Benchmarks}

To compare the different inference algorithms, we use the original benchmarks of ProbZelus~\citep{rppl} as well as two new examples based on realistic applications and one example that is presented as challenging in the original delayed sampling paper~\citep{murray_ds}.
The benchmarks from ProbZelus are the following.

\begin{description}[left= 0pt .. 10pt, labelwidth=*]
\item[Beta-Bernoulli] estimates the bias of a coin from a series of observations. The coin is modeled with a Bernoulli distribution with a Beta prior distribution on the probability of heads.

\item[Gaussian-Gaussian] estimates the mean and variance of a Gaussian distribution from a series of observations. This model has Gaussian priors on the mean and standard deviation.

\item[Kalman-1D] is a one-dimensional Kalman filter that estimates a hidden state from noisy observations.
The state is modeled by a stream of random variables with Gaussian distributions centered on the previous state.

\item[Outlier] is a variation on the Kalman-1D example where the sensor occasionally produces completely invalid observations~\citep{ep}.

\item[Robot] implements a robot controller that computes the commands to reach a target. The controller uses a probabilistic model to estimate the state of the robot (position, velocity, acceleration) using a noisy accelerometer, some sparse noisy GPS observations, and the previous command.

\item[SLAM] (Simultaneous Localization And Mapping) is the problem where a robot has to build a map of an unknown environment in which it travels while estimating its position. 
In the model adapted from~\citet{rbpf}, the agent evolves on a one dimensional discrete black-and-white map where the robot's wheels can slip and the color sensor can produce faulty observations.

\item[MTT] (Multi-Target Tracker) tracks a variable number of moving objects. The model is adapted from~\citet{MurrayS18}. It estimates the path of each object from a set of noisy observations that do not identify the objects.
\end{description}

\medskip

\noindent In addition to these benchmarks, we add the following more challenging models that rely on the closed-family guarantees from Section~\ref{sec:properties} to achieve good performance under \ssdsacro.
By contrast, delayed sampling is unable to maintain exact inference on these benchmarks, and thus falls back on approximate sampling.

\begin{description}[left= 0pt .. 10pt, labelwidth=*]

\item[Tree] is adapted from \citet{lunden17} to illustrate a challenge with delayed sampling. If the symbolic graph forms a binary tree with at least three levels, to observe the leftmost variable and then the rightmost variable, the delayed sampling algorithm samples intermediate nodes and thus fails to produce exact results.

\item[Wheels] is the model presented in Section~\ref{sec:example}.

\item[Delayed GPS] is an extension of the \textit{Robot} benchmark adapted from \citet{delayed_gps} to include a variable delay to the GPS observations. It is also discussed in more detail in Section~\ref{sec:m_path}.
The maximal delay is bounded, so the model can keep a bounded history of the estimated positions to condition the model when a new observation happens.
\end{description}

\subsection{Methodology}

To evaluate the accuracy of the inference algorithms, each benchmark must define an accuracy metric.
Following \citet{rppl}, for the \textit{Robot} and \textit{Delayed GPS} benchmarks, the accuracy metric is the \emph{Linear-Quadratic Regulator}~(LQR) loss~\citep{sontag13control}.
For the \textit{MTT} benchmark, the accuracy metric is a transformation of the \emph{Multiple Object Tracking Accuracy}~\citep{MOTA08} such that it is defined on $[0, \infty)$: $\textrm{MOTA}^* = (1/\textrm{MOTA}) - 1$.
For all the other benchmarks, the accuracy metric is the \emph{Mean Squared Error}~(MSE) of the inferred parameters compared to their exact values.

All the experiments were executed on a server with 64 Intel Xeon E5 CPUs~(2.1GHz) and 128 GB of RAM.
Each benchmark is executed with an increasing number of particles varying from $1$ to $5000$ on a fixed input stream of $500$ time steps.

\begin{figure*}
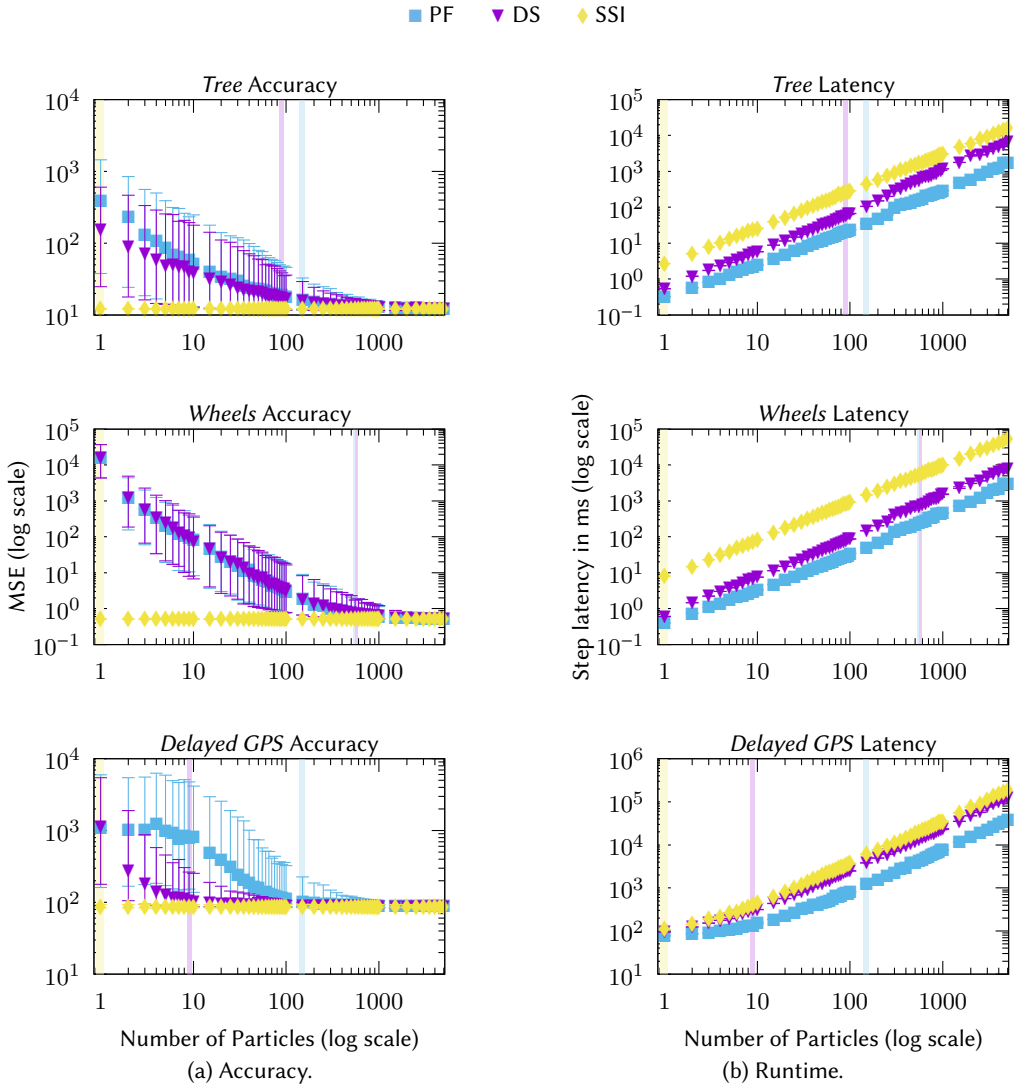

  \begin{center}
    \small\sf
    \pfbullet~PF $\quad$ \pmdsbullet~\gcdsacro $\quad$ \ssdsbullet~\ssdsacro
  \end{center}
    \begin{subfigure}[t]{0.49\textwidth}
      \small\sf
        \hspace*{-1.8em}
        \input{eval_figures/graphs/gtree-accuracy}\\[-0.3cm]
        \hspace*{-1.8em}
        \input{eval_figures/graphs/wheels-accuracy}\\[-0.3cm]
        \hspace*{-1.8em}
        \input{eval_figures/graphs/trackerdelay-accuracy}
        \caption{Accuracy.}
        \label{fig:accuracy3}
    \end{subfigure}
    \hfill
    \begin{subfigure}[t]{0.49\textwidth}
      \small\sf
        \input{eval_figures/graphs/gtree-perf}\\[-0.3cm]
        \input{eval_figures/graphs/wheels-perf}\\[-0.3cm]
        \input{eval_figures/graphs/trackerdelay-perf}
        \caption{Runtime.}
        \label{fig:perf-particles3}
    \end{subfigure}
\caption{Accuracy and runtime as a function of number of particles. The step latency is the delay between two successive time steps. Vertical bars indicates the number of particles where each algorithm reaches the target accuracy reported Table~\ref{table:evaluation}}
\label{fig:eval3}
\vspace{-.25cm}
\end{figure*}

\subsection{Results}

\newcommand{\exact}{\textcolor{green!60!black}{\ding{51} 1}\xspace}
\newcommand{\timeout}{\textcolor{red!80!black}{\ding{55}}\xspace}

\newcommand{\qres}[4]{#1}
\newcommand{\qresuncertainty}[4]{\textcolor{gray}{\tiny{({#2}-{#3})}}}

\begin{table}
\caption{For each benchmark, we report how many particles are required for 90\% of the runs to reach an accuracy close to a target (\gcdsacro with 1000 particles), the corresponding median execution time, and below and in gray the range between 10\% and 90\% execution time quantiles over 1000 runs.  \exact indicates that the algorithm is able to compute the exact solution. \timeout~indicates a timeout.}
\label{table:evaluation}
\vspace{-.25cm}
  \begin{small}
  \begin{tabular}{@{}lrrrrrr@{}}
  &
  \multicolumn{2}{c}{\textsc{PF}} &
  \multicolumn{2}{c}{\textsc{\gcdsacro}} &
  \multicolumn{2}{c}{\textsc{\ssdsacro}}\\
  \cmidrule(lr){2-3}
  \cmidrule(lr){4-5}
  \cmidrule(lr){6-7}
  \textsc{model}
  & \textsc{\# part.} & \textsc{time (ms)}
  & \textsc{\# part.} & \textsc{time (ms)}
  & \textsc{\# part.} & \textsc{time (ms)}
  \\
  \midrule
Beta-Bernoulli & 200 & \qres{23.05}{22.54}{23.90}{1.36} & \exact & \qres{0.28}{0.27}{0.28}{0.01} & \exact & \qres{0.65}{0.65}{0.66}{0.01} \\[-0.4em]
&   & \qresuncertainty{23.05}{22.54}{23.90}{1.36} &   & \qresuncertainty{0.28}{0.27}{0.28}{0.01} &   & \qresuncertainty{0.65}{0.65}{0.66}{0.01} \\
Gaussian-Gaussian & 3000 & \qres{877.44}{689.04}{890.68}{201.64} & 150 & \qres{62.99}{61.69}{65.86}{4.18} & 150 & \qres{182.57}{181.40}{183.54}{2.13} \\[-0.4em]
&   & \qresuncertainty{877.44}{689.04}{890.68}{201.64} &   & \qresuncertainty{62.99}{61.69}{65.86}{4.18} &   & \qresuncertainty{182.57}{181.40}{183.54}{2.13} \\
Kalman-1D & 15 & \qres{3.27}{3.26}{3.28}{0.02} & \exact & \qres{0.33}{0.33}{0.34}{0.01} & \exact & \qres{1.15}{1.14}{1.15}{0.01} \\[-0.4em]
&   & \qresuncertainty{3.27}{3.26}{3.28}{0.02} &   & \qresuncertainty{0.33}{0.33}{0.34}{0.01} &   & \qresuncertainty{1.15}{1.14}{1.15}{0.01} \\
Outlier & 700 & \qres{222.27}{220.76}{223.84}{3.08} & 65 & \qres{43.81}{43.45}{46.48}{3.02} & 65 & \qres{125.88}{125.27}{128.08}{2.82} \\[-0.4em]
&   & \qresuncertainty{222.27}{220.76}{223.84}{3.08} &   & \qresuncertainty{43.81}{43.45}{46.48}{3.02} &   & \qresuncertainty{125.88}{125.27}{128.08}{2.82} \\
Robot & 85 & \qres{771.32}{767.98}{775.51}{7.53} & \exact & \qres{91.44}{90.94}{92.07}{1.13} & \exact & \qres{96.40}{96.21}{97.53}{1.32} \\[-0.4em]
&   & \qresuncertainty{771.32}{767.98}{775.51}{7.53} &   & \qresuncertainty{91.44}{90.94}{92.07}{1.13} &   & \qresuncertainty{96.40}{96.21}{97.53}{1.32} \\
SLAM &   & \timeout & 800 & \qres{2812.55}{2755.99}{2853.89}{97.90} & 800 & \qres{5649.30}{5619.59}{5675.81}{56.23} \\[-0.4em]
&   &   &   & \qresuncertainty{2812.55}{2755.99}{2853.89}{97.90} &   & \qresuncertainty{5649.30}{5619.59}{5675.81}{56.23} \\
MTT &   & \timeout & 60 & \qres{2889.11}{2615.76}{3244.30}{628.55} & 60 & \qres{4457.79}{4068.35}{4996.20}{927.85} \\[-0.4em]
&   &   &   & \qresuncertainty{2889.11}{2615.76}{3244.30}{628.55} &   & \qresuncertainty{4457.79}{4068.35}{4996.20}{927.85} \\
Tree & 150 & \qres{35.55}{35.41}{35.68}{0.26} & 90 & \qres{58.83}{58.55}{59.74}{1.19} & \exact & \qres{2.67}{2.66}{2.70}{0.03} \\[-0.4em]
&   & \qresuncertainty{35.55}{35.41}{35.68}{0.26} &   & \qresuncertainty{58.83}{58.55}{59.74}{1.19} &   & \qresuncertainty{2.67}{2.66}{2.70}{0.03} \\
Wheels & 550 & \qres{246.48}{245.06}{248.75}{3.69} & 550 & \qres{699.12}{672.25}{713.64}{41.38} & \exact & \qres{8.04}{8.00}{8.10}{0.09} \\[-0.4em]
&   & \qresuncertainty{246.48}{245.06}{248.75}{3.69} &   & \qresuncertainty{699.12}{672.25}{713.64}{41.38} &   & \qresuncertainty{8.04}{8.00}{8.10}{0.09} \\
Delayed GPS & 150 & \qres{1221.00}{1218.76}{1230.67}{11.91} & 9 & \qres{304.73}{303.17}{306.31}{3.14} & \exact & \qres{108.55}{108.02}{109.07}{1.04} \\[-0.4em]
&   & \qresuncertainty{1221.00}{1218.76}{1230.67}{11.91} &   & \qresuncertainty{304.73}{303.17}{306.31}{3.14} &   & \qresuncertainty{108.55}{108.02}{109.07}{1.04} \\
\bottomrule
\end{tabular}
\end{small}
\vspace{-.25cm}
\end{table}

Figure~\ref{fig:eval3} presents the results of the evaluation for the \textit{Tree}, \textit{Wheels}, and \textit{Delayed GPS} benchmarks. Figures for the other benchmarks are in Appendix~\ref{sec:graph_appendix}.

Following~\citet{rppl}, to summarize the results in Table~\ref{table:evaluation}, we evaluate how many particles are required for 90\% of the $1000$ runs to reach a target accuracy -- in this case, the median loss of \gcdsacro with 1000 particles: $\log (P_{90\%}(\mathit{loss}))  - \log (\mathit{loss}_{\mathit{target}}) < 0.5$.

\paragraph{RQ1. Accuracy}
Figure~\ref{fig:accuracy3} reports the median accuracy and the $90\%$ and $10\%$ quantiles over $1000$ executions with different random seeds for the \textit{Tree}, \textit{Wheels}, and \textit{Delayed GPS} benchmarks.

Overall, we observe in Table~\ref{table:evaluation} that both \gcdsacro and \ssdsacro outperform PF.
For the three models where \gcdsacro is exact~(\textit{Beta-Bernoulli}, \textit{Kalman-1D}, and \textit{Robot}), \ssdsacro is also able to compute the exact solution.
Moreover, \ssdsacro is exact for the three more challenging models where \gcdsacro requires multiple particles: \textit{Tree}, \textit{Wheels}, and \textit{Delayed GPS}.
For the four remaining models (\textit{Gaussian-Gaussian}, \textit{Outlier}, \textit{SLAM}, and \textit{MTT}), \ssdsacro behaves similarly to \gcdsacro.
More generally, \ssdsacro always outperforms \gcdsacro in accuracy, i.e., \ssdsacro requires the same number of particles or less than \gcdsacro to reach the same accuracy.

\paragraph{RQ2. Speed}
Figure~\ref{fig:perf-particles3} reports the execution time in milliseconds for the three challenging benchmarks and  Table~\ref{table:evaluation} reports the median execution time to reach the target accuracy.

The results show that \ssdsacro is on average $10.5\times$ faster than PF, but there is a
noticeable overhead to run \ssdsacro compared to \gcdsacro, as \ssdsacro is on average $1.6\times$ slower than \gcdsacro.
However, \ssdsacro has a significant speedup for models where \gcdsacro fails to compute the exact solution.
For the more challenging models, compared to \gcdsacro, \ssdsacro is $22\times$ faster on the \textit{Tree} model, $86.9\times$ faster on the \textit{Wheels} model, and $2.8\times$ faster on the \textit{Delayed GPS} model.

\subsection{Discussion: Comparison to Delayed Sampling}
\label{sec:ds_comparison}

In this section we discuss in more detail the reasons why semi-symbolic inference outperforms delayed sampling on the more challenging benchmarks.
In particular, we identify two classes of probabilistic programs on which delayed sampling cannot support exact inference on, but that semi-symbolic inference can: programs with \emph{multiple parents} and programs with \emph{multiple paths}.

\subsubsection{Multiple Parents}

First, we compare delayed sampling and semi-symbolic inference on an example exercising the \emph{multiple parents} case.
This means that this example has a random variable that depends on more than one other random variable in the program (i.e.\ in the symbolic state, the variable will have more than one parent).
This example is the \textit{Wheels} benchmark, presented in full in Section~\ref{sec:example}.
The observed random variables on Lines~\ref{code:obsl} and~\ref{code:obsr} of Figure~\ref{fig:wheels_code} depend on both of the sampled random variables on Lines~\ref{code:omega} and~\ref{code:vel}, and thus each has two parents.

\paragraph{Delayed Sampling}

Delayed sampling does not support random variables with multiple parents~\citep{murray_ds}.
Thus, to execute on this example, delayed sampling must adapt the example to not have multiple parents.
The implementation in ProbZelus performs this adaptation by falling back on approximate sampling.
When ProbZelus executes one of the \zl$observe$ statements and detects that the random variable depends on both of the random variables pointed to by \zl$vel$ and \zl$omega$, it samples either \zl$vel$ or \zl$omega$.
Once this variable has been sampled, the remaining random variables all have at most a single parent, and delayed sampling execution can proceed normally.
However, the sampling step introduces approximation error.

\paragraph{Semi-Symbolic Inference}

This model satisfies the conditions of Theorem~\ref{thm:rbpf} under the linear-Gaussian closed family. Thus, because the program contains no calls to \zl$value$, all symbolic states produced by the algorithm are in the Gaussian closed family and no sampling occurs.

\subsubsection{Multiple Paths}
\label{sec:m_path}

\begin{figure}
\begin{lstlisting}[basicstyle=\small\ttfamily,aboveskip=0em,numbers=left]
let proba model (u, acc, gps) = x where%\label{code:model}%
  rec x = sample (mv_gaussian (last mu, noise))%\label{code:xsample}%
  and init mu = x0 %\label{code:muinit}%
  and mu = (a *@ x) +@ (b *@ u)%\label{code:mu}%
  and () = observe (gaussian (project_2 *@ x, acc_noise), acc)%\label{code:obsacc}%
  and buff_x = buffer(max_delay, x)%\label{code:buffer}%
  and present gps(pos_delay, pos) -> do () = %\label{code:obspos1}%
    observe (gaussian (project_0 *@ get(buff_x, pos_delay), gps_noise), pos) done
\end{lstlisting}
\caption{The Delayed GPS example that exercises the single $m$-path constraint.}
\label{fig:delay_gps}
\vspace{-.35cm}
\end{figure}

One way, originally proposed in \citet{murray_ds}, to circumvent the multiple-parent restriction of delayed sampling is to collapse random variables into \emph{supernodes}.
For example, for the program in Figure~\ref{fig:wheels_code}, the developer could rewrite the example to use multivariate Gaussians and rewrite the arithmetic operators to use matrix multiplication.

However, there exist programs using supernodes that delayed sampling cannot execute purely symbolically.
In this subsection we consider the \textit{Delayed GPS} benchmark that exercises a constraint on delayed sampling called the \emph{single $m$-path constraint}.
Due to this constraint, delayed sampling must also fall back on a sampling-based approximation for this program.

\paragraph{Example}

We consider the example of a robot trying to infer its position using sensors.
The robot's sensors consist of an accelerometer and a GPS receiver.
At each time step, the accelerometer produces a noisy observation of the robot's acceleration.
Intermittently, the robot also receives noisy observations of its position from the GPS receiver.
However, due to delays in the internal processing inherent to any GPS receiver~\citep{delayed_gps}, the GPS signal may be delayed, in that it provides a noisy observation of the position at a previous time step.
The delay can vary, but we assume that it is known as part of the GPS signal.

Figure~\ref{fig:delay_gps} presents how such an example can be encoded in ProbZelus.
Line~\ref{code:model} specifies that the \zl$model$ stream function takes in three parameters: \zl$u$, a stream of \emph{commands} the robot issues to adjust its position; \zl$acc$, a stream of accelerometer inputs; and \zl$gps$, a stream of GPS inputs.
It further specifies that the \zl$model$ stream function returns the hidden state \zl$x$, which is defined by the subsequent set of mutually recursive equations.

Line~\ref{code:xsample} specifies that the hidden state \zl$x$ is sampled from a multivariate Gaussian distribution with mean \zl$last mu$, the previous value of \zl$mu$, and constant covariance matrix \zl$noise$. The hidden state is a vector of length 3 containing numbers that represent the position, velocity, and acceleration of the robot.
Line~\ref{code:muinit} specifies the initial value of \zl$mu$ -- the mean of the hidden state -- to be the constant value~\zl$x0$.
Line~\ref{code:mu} defines \zl$mu$ at subsequent time steps, which is given by a sum of two components.
The first component uses the operator \zl$*@$ -- specifying matrix multiplication -- to multiply the value of the hidden state \zl$x$ with the fixed constant matrix \zl$a$.
The second component multiplies the the input command \zl$u$ with the fixed constant matrix \zl$b$, and the two components are added together with the vector addition operator \zl$+@$.

Line~\ref{code:obsacc} describes the accelerometer process.
It specifies a noisy observation of the component of \zl$x$ at index 2 -- i.e., the robot's acceleration -- which is extracted by multiplying \zl$x$ with the projection matrix given by the constant \zl$project_2$.
The noisy observation consists of a random variable drawn from a Gaussian distribution centered around the projected-out acceleration.
The \zl$observe$ operation conditions the model on this random variable being equal to the input \zl$acc$.

Line~\ref{code:buffer} defines \zl{buff_x}, a sliding window keeping the \zl{max_delay} previous values of \zl{x}, and Line~\ref{code:obspos1} describes the GPS observations.
The \zl$gps$ input stream is a ProbZelus construct called a \emph{signal} that may be present or not at each time step.
The syntax \zl$present gps(pos_delay, pos) ->$ means that whenever the \zl$gps$ has the value -- which is destructed to the pair of \zl$pos_delay$ and \zl$pos$ -- the program will execute whatever follows.
What follows in this case is the observation of the GPS signal.

The observation of the GPS signal uses \zl$get(buff_x, pos_delay)$, which accesses the value of \zl$x$ delayed by \zl$pos_delay$ time steps in the buffered stream \zl$buff_x$.
The program then projects out the position by multiplying with the constant matrix \zl$project_0$.
The observation specifies that the input \zl$pos$ has a Gaussian distribution whose mean is the delayed position and variance is \zl$gps_noise$.

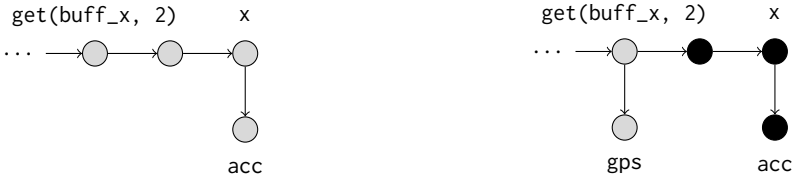
\begin{figure}
\begin{subfigure}[t]{0.45\textwidth}
\centering
\begin{tikzpicture}
\draw (-1, 0) node (predots) {$\dots$};

\draw (0, 0) node [draw=black, fill=gray!30, circle] (node1) {};
\draw (0, 0.5) node (delaylabel) {\zl$get(buff_x, 2)$};

\draw (1, 0) node [draw=black, fill=gray!30, circle] (node2) {};

\draw (2, 0) node [draw=black, fill=gray!30, circle] (node3) {};
\draw (2, 0.5) node (xlabel) {\zl$x$};

\draw (2, -1) node [draw=black, fill=gray!30, circle] (accnode) {};
\draw (2, -1.5) node (acclabel) {\zl$acc$};

\draw[->] (predots) -- (node1);
\draw[->] (node1) -- (node2);
\draw[->] (node2) -- (node3);
\draw[->] (node3) -- (accnode);
\end{tikzpicture}
\caption{ The delayed sampling symbolic state after the inference system observes the acceleration \zl$acc$ all nodes must be marginalized at this point.}
\end{subfigure}$\qquad$
\begin{subfigure}[t]{0.45\textwidth}
\centering
\begin{tikzpicture}
\draw (-1, 0) node (predots) {$\dots$};

\draw (0, 0) node [draw=black, fill=gray!30, circle] (node1) {};
\draw (0, 0.5) node (delaylabel) {\zl$get(buff_x, 2)$};

\draw (0, -1) node [draw=black, fill=gray!30, circle] (gpsnode) {};
\draw (0, -1.5) node (gpslabel) {\zl$gps$};

\draw (1, 0) node [draw=black, fill=black, circle] (node2) {};

\draw (2, 0) node [draw=black, fill=black, circle] (node3) {};
\draw (2, 0.5) node (xlabel) {\zl$x$};

\draw (2, -1) node [draw=black, fill=black, circle] (accnode) {};
\draw (2, -1.5) node (acclabel) {\zl$acc$};

\draw[->] (predots) -- (node1);
\draw[->] (node1) -- (node2);
\draw[->] (node2) -- (node3);
\draw[->] (node3) -- (accnode);
\draw[->] (node1) -- (gpsnode);
\end{tikzpicture}
\caption{ The delayed sampling state after the inference system observes the delayed GPS signal \zl$gps$. Two of the hidden states must become realized in order to satisfy the single $m$-path constraint.}
\end{subfigure}

\caption{A depiction of how delayed sampling executes on the program in Figure~\ref{fig:delay_gps}. We depict marginalized nodes in gray and realized nodes in black.
Arrows mean that a given random variable depends on another random variable (i.e., is sampled or observed from other variable).
We have also labeled the random variables pointed to by the current hidden state \zl$x$, the 2-step delayed hidden state \zl$get(buff_x, 2)$, the observed acceleration \zl$acc$, and the observed GPS signal \zl$gps$.}
\label{fig:ds_diag}
\vspace{-.35cm}
\end{figure}

\paragraph{Delayed Sampling}

This example exercises a constraint of delayed sampling called the \emph{single $m$-path constraint}.
In delayed sampling, random variables are represented using one of three types of nodes in a graph: initialized, marginalized, or realized.
Over the course of a delayed sampling execution, nodes change state from initialized to marginalized and from marginalized to realized.
Notably, realizing a variable samples from a distribution and thus loses accuracy.

In delayed sampling, the symbolic state may only contain a single $m$-path: a path from the root containing all marginalized nodes.\footnote{ In general, delayed sampling can have multiple $m$-paths, but may only have one $m$-path per tree in the graph. Our example has only one tree, so may have only one $m$-path.}
Furthermore, any variable that is being \zl$observe$d and all of its ancestors must be marginalized.
Thus, in the example in Figure~\ref{fig:delay_gps}, when the delayed GPS signal is observed after the acceleration, the inference system moves the $m$-path from the acceleration variable to the delayed GPS variable, and all marginalized hidden states in between the current and delayed time steps are realized to preserve the single $m$-path constraint.
Figure~\ref{fig:ds_diag} depicts this case.

\paragraph{Semi-Symbolic Inference}

In our implementation of semi-symbolic inference in ProbZelus, we have implemented rules for multivariate Gaussians similar to the rules for the univariate Gaussians in Algorithm~\ref{alg:swap}.
Multivariate Gaussians also form a closed family, and this model always produces symbolic states in that closed family.
Thus, the model will execute fully symbolically and the runtime inference system will not draw any samples.
The underlying reason is that semi-symbolic inference's swaps are fully reversible transformations.
By contrast, when delayed sampling changes a node's state, it cannot change it back, and sets the execution on an irreversible path towards sampling-based approximation.

\section{Related Work}

In this section, we compare semi-symbolic inference to various related approaches.

\subsection{Exact Inference Systems}

Some probabilistic programming systems are designed specifically for exact inference.
Examples include the PSI Solver~\citep{psi}, Dice~\citep{dice}, SPPL~\citep{sppl}, and Autoconj~\cite{autoconj}.
Some of these languages also take advantage of closed families.
For example, Dice focuses on exact inference with problems that only have finite discrete random variables.
Exact inference is always possible on these models, and Dice thus focuses on improving the computational efficiency of exact inference.
Other systems, such as PSI, use a complex solver to potentially solve a much larger class of programs.
The Autoconj system analyzes probabilistic programs using a similar symbolic representation to that presented in Section~\ref{sec:semisymb}.
In general, these systems do not support the symbolic interface we discuss in Section~\ref{sec:semisymb_sec}, inhibiting our ability to use them to build delayed sampling runtime inference systems.

\subsection{Delayed Sampling}

We compared this work against ProbZelus's delayed sampling system because prior work on ProbZelus~\citep{rppl} established that delayed sampling is an effective technique for implementing RBPFs within the inference system of a streaming probabilistic programming language.
However, as we discuss in Section~\ref{sec:ds_comparison}, the limitations of ProbZelus's delayed sampling system means that is not able to provide exact inference in all cases that semi-symbolic inference can.

ProbZelus uses the same delayed sampling system as Birch~\citep{murray_ds} and Anglican~\citep{lunden17}, and we expect these limitations to apply to these other languages as well.
Furthermore, because Birch and Anglican expose the same symbolic interface as ProbZelus, we anticipate semi-symbolic inference could be applied to these languages.

Pyro~\citep{pyro} supports delayed sampling using an alternative symbolic representation called functional tensors or \emph{funsors}~\citep{funsors}.
Pyro performs exact inference on funsors using an alternative symbolic interface based on \emph{variable elimination}~\cite{ve,tove}.
Variable elimination works by removing variables from the symbolic state.
For batch execution of probabilistic programs, all remaining variables at the end of the execution can be eliminated, but in a streaming context this approach needs to decide when to eliminate variables.
This decision needs to balance 1) eliminating old variables to limit the size of the symbolic state, 2) eliminating variables to compute particle weights, and 3) keeping variables available for use at future time steps.
Due to these tradeoffs, more work is needed to develop a delayed sampling system for streaming probabilistic programs based on variable elimination.

\subsection{Alternatives for Combining Exact and Approximate Inference}

\paragraph{Hakaru.} Hakaru~\citep{hakaru} is a probabilistic programming system that statically rewrites probabilistic programs into inference procedures.
This includes transformations that introduce sampling-based approximations as well as analytically solving some distributions with a solver.
One tradeoff between Hakaru and this work is the inherent tradeoff between static and dynamic techniques.
Hakaru relies on a sufficiently powerful static analysis to determine if exact inference is possible, whereas semi-symbolic inference will exploit the closed-family guarantees so long as the program satisfies the necessary conditions at runtime.

\paragraph{Stochastic Procedures.} Languages such as Venture~\citep{venture} support encapsulating exact inference components inside \emph{stochastic procedure} objects.
These stochastic procedures can then be used inside of various approximate inference algorithms.
However, these require developers to rewrite their input probabilistic models to use stochastic procedures, which breaks the separation between modeling and inference.
By contrast, semi-symbolic inference exists as an alternative inference technique inside the language runtime inference system and only requires developers to add calls to \zl$value$ to control where sampling occurs.

\paragraph{Shared Variables.}
Infer.NET~\cite{infernet} provides language support for inference on graphical models, including exact inference with belief propagation~\cite{belief-propagation}.
Through its feature of \emph{shared variables}, Infer.NET supports combining belief propagation with the approximate inference algorithms of expectation propagation~\cite{ep}, variational message passing~\cite{vmp}, and Gibbs sampling~\cite{gibbs}.
This does not present the same interface as Definition~\ref{def:interface}.
Instead, it provides an alternative way of combining exact and approximate inference that does not result in an RBPF inference algorithm.

\section{Conclusion}
In this paper, we presented semi-symbolic inference, a novel technique for combining exact and approximate probabilistic inference.
It enables developers to write models in a high-level streaming probabilistic programming language while the language runtime inference system automatically implements Rao-Blackwellized particle filtering.
It presents developers of streaming probabilistic programs with the opportunity to combine the efficiency of exact inference with the generality of sampling-based approximate inference to achieve the performance they expect.


\begin{acks}
This work was supported in part by the MIT-IBM Watson AI Lab and the Office of Naval Research (ONR N00014-17-1-2699).
Any opinions, findings, and conclusions or recommendations expressed in this material are those of the author and do not necessarily reflect the views of the Office of Naval Research.
\end{acks}

\bibliography{biblio}

\appendix
\section{Detailed Semantics and Proofs}
\label{sec:proof_app}

\subsection{Semantics of Symbolic State}
\label{sec:symb_state_sem}

In this section, we describe the meaning of the constructs in Figure~\ref{fig:symb-grammar}.
Note that this is not the same as the semantics of ProbZelus itself as described in Section~\ref{sec:background}.

\newcommand{\semenv}{\sigma}
\newcommand{\Semenv}{\Sigma}

\paragraph{Denotation of Expressions}
The denotation of an expression $e$ is $\sem{e}$, which takes in an environment that maps random variable names to values and returns a value.
$\sem{e}$ produces either a real number or integer.
It is defined as follows, with remaining analogous cases elided:
\begin{align*}
\sem{\randomvar}(\semenv) & = \semenv(\randomvar)\\
\sem{\real}(\semenv) & = \real \\
\sem{\plus{\expr_1}{\expr_2}}(\semenv) & = \sem{e_1}(\semenv) + \sem{e_2}(\semenv)\\
\sem{\minus{\expr_1}{\expr_2}}(\semenv) & = \dots
\end{align*}

\paragraph{Denotation of Distributions}
The denotation of an individual distribution $\distr$ is $\sem{\distr}$, which returns the measure the distribution is associated with.
The measure is either over real numbers or integers, depending on the type of the distribution.
\begin{align*}
\sem{\normal{\expr_1}{\expr_2}}(\semenv) & = \mu_{\mathrm{Gaussian}}(\sem{\expr_1}(\semenv), \sem{\expr_2}(\semenv))\\
\sem{\betad{\expr_1}{\expr_2}}(\semenv) & = \mu_{\mathrm{Beta}}(\sem{\expr_1}(\semenv), \sem{\expr_2}(\semenv))\\
\sem{\bern{\expr}}(\semenv) & = \mu_{\mathrm{Bernoulli}}(\sem{\expr}(\semenv))\\
\sem{\deltad{\expr}}(\semenv) & = \delta(\sem{\expr}(\semenv))
\end{align*}
where $\mu_{\mathrm{Gaussian}}$, $\mu_{\mathrm{Beta}}$, and $\mu_{\mathrm{Bernoulli}}$ are the standard measures for Gaussian, Beta, and Bernoulli distributions, respectively.
We also use, by abuse of notation, $\delta$ to refer to the symbolic Delta distribution as well as its measure interpretation.

\paragraph{Denotation of Symbolic States}
The denotation of a symbolic state $\pstate$ is $\sem{\pstate}$, which is a joint distribution over the random variables $\mathcal{V}$ mapped in $\pstate$.
We use the notation $\randomvar_1, \randomvar_2, \dots, \randomvar_n$ to refer to a fixed order of the random variables in the joint distribution.

We now define what $\sem{\pstate}$ is, and our general approach is as follows.
We first establish an ordering between the random variables that is consistent with their dependencies.
Then, we proceed through the variables in this order, using $\sem{\pstate}^i$ to refer the intermediate semantics of $\pstate$ after incorporating the $i$th variable.
The definition of $\sem{\pstate}^i$ is defined by using the probability monad~\cite{giry} to specify the measure of the $i$th variable given the measures of all variables it depends on (i.e.\ all variables that precede $i$ in the ordering).
Then, after incorporating all variables, we ultimately define $\sem{\pstate}$ from $\sem{\pstate}^n$ by remapping variables from the dependency-consistent order back into the original order in the product space.

Let $\randomvar_{(1)}, \randomvar_{(2)}, \dots, \randomvar_{(n)}$ be an ordering of the elements in $\mathcal{V}$ that is consistent with the dependencies of $\mathcal{V}$, i.e.\ if $\randomvar_j$ depends on $\randomvar_i$, then $i < j$.
We can then define the intermediate semantics $\sem{\pstate}^i$ as follows.
This definition returns a joint distribution over the first $i$ variables of the ordering, and a permutation in $\Randomvar \rightarrow \Randomvar$ mapping each variable back into the original product space.
\begin{align*}
\sem{\pstate}^0 ={} & \delta(())\\
\sem{\pstate}^i ={} & \mathit{let} \; (d, \pi) = \sem{\pstate}^{i-1} \; \mathit{in} \\
                  & (d \bind \lambda (x_1, x_2, \dots, x_{i-1}). \; \sem{\pstate(\randomvar_{(i)})}[\randomvar_{(1)} \mapsto x_1, \randomvar_{(2)} \mapsto x_2, \dots, \randomvar_{(i-1)} \mapsto x_{i-1}] \bind \\
                  & \lambda x_i. \; \mathbf{return} \; (x_1, x_2, \dots, x_i)  , \pi[\randomvar_{(i)} \mapsto \randomvar_i])
\end{align*}
Here, the $\bind$ and $\mathbf{return}$ operators are the bind and return operators of the probability monad~\cite{giry}.
We can then define the semantics of $\pstate$ as the pushforward of the last intermediate semantics across the finite mapping:
$$
\sem{\pstate} = \mathit{let} \; (d, \pi) = \sem{\pstate}^n \; \mathit{in} \; d_*\pi
$$

\paragraph{Conditional Distributions}
Given a joint distribution $\Pr_g(\mathcal{V}) = \sem{\pstate}$, we define the conditional distribution $\Pr_g(\mathcal{V}' \mid \randomvar_1 = x_1, \randomvar_2 = x_2, \dots, \randomvar_n = x_n)$, where $\mathcal{V}'=\mathcal{V} \setminus \{\randomvar_1, \randomvar_2, \dots, \randomvar_n\}$, using disintegration~\cite{disintegration}.
This disintegration maintains the following properties.
Note that disintegrations are uniquely defined up to a measure of size zero; thus in the following formalism, equality between disintegrations should be taken to mean almost-sure equality.
\begin{itemize}
    \item {\it Marginalization.} If $\mathcal{V}' \subset \mathcal{V}$ then the distribution $\Pr_g(\mathcal{V'}) = \int_{\hat{x} \in \mathcal{V} \setminus \mathcal{V}'} \Pr_g(\mathcal{V})$.
    \item {\it Transitivity of Conditioning.} If $\Pr_g(\mathcal{V}) = \Pr_{\pstate'}(\mathcal{V} \mid \randomvar_1 = x_1)$ and we have that $\Pr_{\pstate'}(\mathcal{V}, \randomvar_1) = \Pr_{\pstate''}(\mathcal{V}, \randomvar_1 \mid \randomvar_2 = x_2)$, then $\Pr_g(\mathcal{V}) = \Pr_{\pstate''}(\mathcal{V} \mid \randomvar_1 = x_1, \randomvar_2 = x_2)$.
    \item {\it Bayes' Rule.} If $\mathcal{V}' = \mathcal{V} \setminus \{\randomvar_1, \randomvar_2. \dots, \randomvar_n\}$, and $\Pr_g(\mathcal{V})$ and $\Pr_g(\{ \randomvar_1, \randomvar_2, \dots, \randomvar_n \})$ have densities\footnote{These densities are with respect to a natural combination of the counting and Lebesgue measures depending on whether each random variable is an integer or real random variable, respectively. } $p(\hat{\randomvar})$ and $p'(\hat{\randomvar})$, then the distribution $\Pr_g(\mathcal{V}' \mid \randomvar_1 = x_1, \randomvar_2 = x_2, \dots, \randomvar_n = x_n)$ has the density $p_{cond}(\hat{\randomvar}) = \frac{p(\hat{\randomvar})}{p'(\hat{\randomvar})}$.
    \item {\it Probability Monad.} If $\mathcal{V'} = \{\randomvar_1, \randomvar_2. \dots, \randomvar_n\}$ is a subset of $\mathcal{V}$, then $\Pr_g(\mathcal{V}) = \Pr_g(\mathcal{V} \setminus \mathcal{V'}) \bind \lambda (x_1, x_2, \dots, x_n). \; \Pr_g(\mathcal{V} \mid \randomvar_1 = x_1, \randomvar_2 = x_2, \dots, \randomvar_n = x_n)$.
\end{itemize}

\subsection{Correctness Proofs}
\label{sec:semisymb_correctness_proofs}

The following axiom specifies the behavior of the \textsc{draw} and \textsc{score} primitives.

\begin{axiom}[Draw and Score]
If a random variable $\randomvar$ passed into \textsc{draw} or \textsc{score} is a root, then \textsc{draw} draws a sample from its distribution and \textsc{score} evaluates its probability density.
\label{ax:draw_score}
\end{axiom}

The following lemma states that a \textsc{swap} operation preserves the denotation of the symbolic state.

\newtheorem*{swaplem}{Lemma~\ref{lem:swap_preservation}}
\begin{swaplem}[Swap Preservation]
If $(\pstate', \_) = \textsc{swap}(\randomvar_1, \randomvar_2, \pstate)$, then $\sem{\pstate'} = \sem{\pstate}$.
\end{swaplem}
\begin{proof}
We consider two cases, based on whether or not \textsc{swap} is able to find a conjugate relationship.
If not, then $\pstate' = \pstate$ and the theorem holds by definition.

Otherwise, let $d_1 = \Pr_g(\randomvar_1 \mid \randomvar_<)$ and $d_2 = \Pr_g(\randomvar_2 \mid \randomvar_1, \randomvar_<)$ where $\randomvar_<$ is the set of random variables that come before $\randomvar_1$ in the ordering for $\mathcal{V}$ that defines the semantics of $\pstate$, and $d_1$ and $d_2$ are uniquely specified by the semantics of $\pstate$ and the probability monad property.
Under the assumption that \textsc{swap} is able to find a conjugacy relationship, both $d_1$ and $d_2$ will have valid densities.
Furthermore, we can obtain densities for $d_2' = \Pr_g(\randomvar_2 \mid \randomvar_<)$ and $d_1' = \Pr_g(\randomvar_1 \mid \randomvar_2, \randomvar_<)$ using the marginalization and Bayes' rule properties.
According to the rules of conjugate priors~\citep{conjugate_priors}, the returned distributions of \textsc{swap} have densities that are equal to $d_2'$ and $d_1'$.
Letting $k$ be such that $\randomvar_{(k)} = \randomvar_2$ in $\pstate'$, because in $\pstate'$ $\randomvar_2$ depends on $\randomvar_1$, $\sem{\pstate'}^{k}$ therefore represents the distribution of $\Pr_\pstate(\randomvar_1, \randomvar_2, \randomvar_<)$.
Because the distributions of the remaining variables are unchanged, this means that $\sem{\pstate'} = \Pr_\pstate(\Randomvar) = \sem{\pstate}$.
\end{proof}

The following set of lemmas formalize the correct behavior of the helper functions \textsc{eval}, \textsc{hoist\_helper}, \textsc{intervene}, \textsc{hoist}, \textsc{value}, and \textsc{observe}.

\begin{lemma}[\textsc{eval} Correctness]
We say that $\pstate \vDash \semenv$ iff $\forall \randomvar.\; \pstate(\randomvar) = \delta(\expr) \Rightarrow \semenv(\randomvar) = \sem{e}(\semenv)$.

If $\expr'$ = $\textsc{eval}(\expr, \pstate)$ and $\pstate \vDash \semenv$, then $\sem{\expr'}(\semenv) = \sem{\expr}(\semenv)$.
\label{lem:eval}
\end{lemma}
\begin{proof}
By structural induction on expressions.
\end{proof}

\newtheorem*{lemevalstar}{Lemma~\ref{lem:eval_star}}
\begin{lemevalstar}[$\textsc{eval}^*$ Correctness]
If $\pstate' = \textsc{eval}^*(\randomvar, \pstate)$, then $\sem{\pstate'} = \sem{\pstate}$.
\end{lemevalstar}
\begin{proof}
We apply the definitions of $\sem{\pstate}$, $\sem{\pstate'}$, and Lemma~\ref{lem:eval}.
We further apply the fact that for any $\randomvar_i$, if $\pstate(\randomvar_i) = \delta(\expr)$, then the corresponding value $x_i$ drawn in the definitions of $\sem{\pstate}$ and $\sem{\pstate'}$ will be equal to the value of the expression $\expr$.
\end{proof}

\newtheorem*{lemhh}{Lemma~\ref{lem:hoist_helper}}
\begin{lemhh}[\textsc{hoist\_helper} Correctness]
If $\pstate' = \textsc{hoist\_helper}(\randomvar_\mathrm{cur}, \texttt{roots}, \pstate)$, and no exceptions are thrown, then $\sem{\pstate'} = \sem{\pstate}$ and in $\pstate'$, $\randomvar_\mathrm{cur}$ is a root, except it may depend on variables in \texttt{roots}.
\end{lemhh}
\begin{proof}
We proceed by induction on executions of \textsc{hoist\_helper}.
At the end of the recursive calls, by inductive hypothesis, the input variable has only immediate parents (except for variables originally in \texttt{roots}).
We then swap with all roots in the swap phase, meaning that any remaining ancestors must be in \texttt{roots}.

The preservation of the denotation of the symbolic state follows from the fact that all symbolic states are constructed from the original using the \textsc{swap} function, and so we can apply Lemma~\ref{lem:swap_preservation}.
\end{proof}

\begin{lemma}[\textsc{intervene} Correctness]
If $\randomvar$ is a root in $\pstate$, and $\pstate' = \textsc{intervene}(\randomvar, r, \pstate)$, then for any subset $\mathcal{V}'$ of random variables mapped in $\pstate$, $\Pr_{\pstate'}(\mathcal{V}') = \Pr_{\pstate}(\mathcal{V}' \mid \randomvar = r)$.
\label{lem:intervention}
\end{lemma}

\newtheorem*{lemintervene}{Lemma~\ref{lem:intervention}}
\begin{lemintervene}[\textsc{intervene} Correctness]
If $\randomvar$ is a root in $\pstate$, and $\pstate' = \textsc{intervene}(\randomvar, r, \pstate)$, then for any subset $\mathcal{V}$ of random variables mapped in $\pstate$, $\Pr_{\pstate'}(\mathcal{V}) = \Pr_{\pstate}(\mathcal{V} \mid \randomvar = r)$.
\end{lemintervene}
\begin{proof}
Apply the semantics of $\pstate$ and $\pstate'$ and the probability monad property.
\end{proof}

\newtheorem*{thmhoist}{Theorem~\ref{thm:hoist_value}}
\begin{thmhoist}[\textsc{hoist} and \textsc{value} Correctness]
If $\pstate' = \textsc{hoist}(\randomvar_\mathrm{in}, \pstate)$, then $\Pr_{\pstate'}(\Randomvar) = \Pr_\pstate(\Randomvar \mid \hat{V} = \hat{v})$, where $\hat{V}$ is the set of variables sampled during the execution of \textsc{hoist} and $\hat{v}$ is the corresponding set of sampled values.
Furthermore, after executing \textsc{hoist}, $\randomvar_\mathrm{in}$ is a root in $\pstate'$.

Also, if $(\pstate', v) = \textsc{value}(\randomvar, \pstate)$, then $v$ is a sample from $\Pr_g(\randomvar \mid \hat{V} = \hat{v})$ and $\Pr_{\pstate'}(\Randomvar) = \Pr_g(\Randomvar \mid \randomvar = v, \hat{V} = \hat{v})$, where $\hat{V}$ and $\hat{v}$ are as above.
\end{thmhoist}
\begin{proof}
We proceed by induction on the mutual recursion between \textsc{hoist} and \textsc{value}.

For \textsc{hoist}, there are two cases: one where \textsc{hoist\_helper} executes without error, in which case we simply apply~\ref{lem:hoist_helper}, and one where it does throw an error.
In this second case, we apply Lemma~\ref{lem:eval_star} and the inductive hypothesis for \textsc{value}.

For \textsc{value}, we apply the inductive hypothesis for \textsc{hoist}, Axiom~\ref{ax:draw_score}, Lemma~\ref{lem:eval_star} and Lemma~\ref{lem:intervention}.
\end{proof}

\newtheorem*{thmobs}{Theorem~\ref{thm:observe}}
\begin{thmobs}[\textsc{observe} Correctness]
If $(\pstate', w) = \textsc{observe}(\randomvar, r, \pstate)$, then we have that $\Pr_{\pstate'}(\Randomvar) = \Pr_{\pstate}(\Randomvar \mid \randomvar = r, \hat{V} = \hat{v})$, where $\hat{V}$ and $\hat{v}$ are the random variables that may need to be sampled during the observation, and $\hat{v}$ are their sampled values.
Furthermore, $w$ is the density of $\Pr_{\pstate}(\randomvar \mid \hat{V} = \hat{v})$.
\end{thmobs}
\begin{proof}
Apply Axiom~\ref{ax:draw_score}, Theorem~\ref{thm:hoist_value}, Lemma~\ref{lem:eval_star} and Lemma~\ref{lem:intervention}.
\end{proof}

We next define the correctness of the overall symbolic interface. To do so, we first define the concept of an \emph{execution}. An execution is any sequence of symbolic states that is reachable by composing functions in the symbolic interface.

\begin{definition}[Execution]
An execution is a sequence of symbolic states $(\pstate_i)$, where $i \in \mathbb{N}$, such that $\pstate_0 = \emptyset$ and for all $i > 0$, either:
\begin{itemize}
\item $\pstate_i$ = $\textsc{assume}(\mu, \pstate_{i-1})$ for some $\mu$,
\item $(\pstate_i, v)$ = $\textsc{value}(\randomvar, \pstate_{i-1})$ for some $\randomvar$ and $v$, or
\item $(\pstate_i, w)$ = $\textsc{observe}(\randomvar, v, \pstate_{i-1})$ for some $\randomvar$, $v$, and $w$.
\end{itemize}
\end{definition}

We also define the concept of an \emph{erased execution} which, given an original execution, retains all the \textsc{assume} operations in the original but discards all other operations.

\begin{definition}[Erased Execution]
Given an execution $(\pstate_i)$ for $i \in \mathbb{N}$, an erased execution $(\pstate_i^*)$ for $i \in \mathbb{N}$ is defined as follows. We define that $\pstate_0^* = \emptyset$, and for $i > 0$,
\begin{itemize}
\item if $\pstate_i$ = $\textsc{assume}(\mu, \pstate_{i-1})$ for some $\mu$, then $\pstate_i^*$ = $\textsc{assume}(\mu, \pstate_{i-1}^*)$, or
\item otherwise, $\pstate_i^* = \pstate_{i-1}^*$.
\end{itemize}
\end{definition}

We now formalize the correctness of the semi-symbolic interface. This theorem states that the symbolic state represents the joint distribution of all random variables conditioned on the values of all variables that have been passed to \textsc{value} and \textsc{observe}. The overall joint distribution is defined using erased executions that retain the information from all \textsc{assume} operations.

Let $\pstate_i$ be a symbolic state that is the $i$th element of some execution. Let $\hat{V}$ and $\hat{v}$ be all previously \textsc{value}d random variables and their corresponding random samples in the execution that produced $\pstate_i$. Similarly, let $\hat{O}$ and $\hat{o}$ be all previously \textsc{observe}d random variables in the execution producing $g_i$ and their corresponding observed values.

\begin{theorem}[Semi-Symbolic Interface Correctness]
For any subset $\mathcal{V}'$ of the random variables mapped in $\pstate_i$, $\Pr_{\pstate_i}(\mathcal{V}') = \Pr_{\pstate_i^*}(\mathcal{V}' \mid \hat{V} = \hat{v}, \hat{O} = \hat{o})$.
Also, if $(\pstate_i, v)$ = $\textsc{value}(\randomvar, \pstate_{i-1})$, then $v$ is sampled from $\Pr_{\pstate_{i-1}}(\randomvar)$. Furthermore, if $(\pstate_i, w)$ = $\textsc{observe}(\randomvar, v, \pstate_{i-1})$, then $w$ is the value of the probability density for the distribution of $\Pr_{\pstate_{i-1}}(\randomvar)$.
\label{thm:correctness}
\end{theorem}
\begin{proof}
We proceed by induction on $i$. In the base case, $i = 0$, we know that $\mathcal{V} = \emptyset$ and the theorem holds trivially.

For the inductive step, we proceed by cases depending on how $\pstate_i$ is constructed from $\pstate_{i-1}$. In the case of \textsc{assume}, we use the inductive hypothesis and the fact that according to the definitions, \textsc{assume} is deterministic. For \textsc{value}, we apply Theorem~\ref{thm:hoist_value} and the transitivity of conditioning property. For \textsc{observe}, we apply Theorem~\ref{thm:observe} and the transitivity of conditioning property.
\end{proof}

\subsection{Additional Side Conditions}
\label{sec:proof_side_conditions}

In this section, we prove additional side conditions that ensure the main theorems above hold. First, we show that the \textsc{can\_swap} assertion will always succeed.

\begin{theorem}[\textsc{can\_swap} succeeds] \label{thm:can-swap-succeeds}
In any execution of \textsc{hoist\_helper}, $\textsc{can\_swap}(\randomvar_\mathrm{par}, \randomvar_\mathrm{cur}, \pstate)$ succeeds.
\end{theorem}
\begin{proof}
Because during this part of \textsc{hoist\_helper}, we are iterating in reverse topological sort order, the loop maintains the invariant that the current parent $\randomvar_\mathrm{par}$ does not have $\randomvar_\mathrm{cur}$ as an ancestor.
This is because the only ancestors of $\randomvar_\mathrm{par}$ are parents that have previously been visited on an earlier iteration, and thus have already been swapped with $\randomvar_\mathrm{cur}$ and cannot have $\randomvar_\mathrm{cur}$ as a parent.
\end{proof}

\paragraph{Hoist Termination}

Next, we show that \textsc{hoist\_helper} and thus \textsc{hoist} terminate.

\begin{lemma}[\textsc{hoist\_helper} termination] \label{lem:helper-term}
Starting from any valid (i.e.\ acyclic) symbolic state $\pstate$, any $\randomvar$ in the domain of $\pstate$, and \texttt{roots}, a subset of the domain of $\pstate$, $\textsc{hoist\_helper}(\randomvar, \texttt{roots})$ terminates.
\end{lemma}
\begin{proof}
We use the following strictly decreasing metric on successive calls to \textsc{hoist\_helper}: the number of ancestors of the input variable, \emph{except} ancestors that are only reachable through the \texttt{roots} set.
\end{proof}

Now, we use this result to show that \textsc{hoist} terminates overall:
\begin{theorem}[\textsc{hoist} termination]
The mutual recursion between $\textsc{hoist}(\randomvar, \pstate)$ and $\textsc{value}(\randomvar, \pstate)$ terminates for any valid (i.e.\ acyclic) symbolic state $\pstate$ and any random variable $\randomvar$ in the domain of $\pstate$.
\end{theorem}
\begin{proof}
We use the following decreasing method on successive calls to \textsc{hoist}: the lexicographic ordering combining a) the number of nodes that are \emph{not} a $\delta$ distribution, and b) the number of ancestors of the input node.

The proof proceeds by cases:
\begin{itemize}
\item {\em Case 1.} No exception is thrown. In this case, we apply Lemma~\ref{lem:helper-term}.
\item {\em Case 2.} An exception is thrown and caught by the \textbf{catch} block.
In this case, the thrown parent $\randomvar_\mathrm{par}$ is a strict ancestor if the input variable $\randomvar_\mathrm{in}$, and ancestors of $\randomvar_\mathrm{par}$ may not include $\randomvar_\mathrm{in}$, otherwise the $\textsc{can\_swap}$ assertion would fail.
Thus, \textsc{value} will be called with a variable that has a strict subset of ancestors of the input variable, and will recurse back into \textsc{hoist} with this variable.
Once this recursion is complete, the tail recursive call to \textsc{hoist} will have an additional $\delta$ distribution.
\end{itemize}
\end{proof}

\section{Additional Performance Evaluation Figures}
\label{sec:graph_appendix}

\begin{figure*}[p]
  \begin{center}
    \small\sf
    \pfbullet~PF $\quad$ \pmdsbullet~\gcdsacro $\quad$ \ssdsbullet~\ssdsacro
  \end{center}
    \begin{minipage}[t]{0.49\textwidth}
      \small\sf
        \hspace*{-1.8em}
        \input{eval_figures/graphs/coin-accuracy}\\[-0.3cm]
        \hspace*{-1.8em}
        \input{eval_figures/graphs/gaussian-accuracy}\\[-0.3cm]
        \hspace*{-1.8em}
        \input{eval_figures/graphs/kalman-accuracy}\\[-0.3cm]
        \hspace*{-1.8em}
        \input{eval_figures/graphs/outlier-accuracy}
        \caption{Accuracy as a function of the number of particles.}
        \label{fig:accuracy1}
    \end{minipage}
    \hfill
    \begin{minipage}[t]{0.49\textwidth}
      \small\sf
        \hspace*{-1.8em}
        \input{eval_figures/graphs/coin-perf}\\[-0.3cm]
        \hspace*{-1.8em}
        \input{eval_figures/graphs/gaussian-perf}\\[-0.3cm]
        \hspace*{-1.8em}
        \input{eval_figures/graphs/kalman-perf}\\[-0.3cm]
        \hspace*{-1.8em}
        \input{eval_figures/graphs/outlier-perf}
        \caption{Runtime performance as a function of particles.}
        \label{fig:perf-particles1}
    \end{minipage}
\end{figure*}

\begin{figure*}[p]
  \begin{center}
    \small\sf
    \pfbullet~PF $\quad$ \pmdsbullet~\gcdsacro $\quad$ \ssdsbullet~\ssdsacro
  \end{center}
    \begin{minipage}[t]{0.49\textwidth}
      \small\sf
        \hspace*{-1.8em}
        \input{eval_figures/graphs/tracker-accuracy}\\[-0.3cm]
        \hspace*{-1.8em}
        \input{eval_figures/graphs/slam-accuracy}\\[-0.3cm]
        \hspace*{-1.8em}
        \input{eval_figures/graphs/mtt-accuracy}
        \caption{Accuracy as a function of the number of particles.}
        \label{fig:accuracy2}
    \end{minipage}
    \hfill
    \begin{minipage}[t]{0.49\textwidth}
      \small\sf
        \input{eval_figures/graphs/tracker-perf}\\[-0.3cm]
        \input{eval_figures/graphs/slam-perf}\\[-0.3cm]
        \input{eval_figures/graphs/mtt-perf}
        \caption{Runtime performance as a function of particles.}
        \label{fig:perf-particles2}
    \end{minipage}
\end{figure*}

\end{document}